\documentclass[usenatbib]{mn2e}
\usepackage{graphicx}
\usepackage{ifthen}
\usepackage{amsmath}
\usepackage{url}
\usepackage{rotating}

\def\ltsima{$\; \buildrel < \over \sim \;$}
\def\lta{\lower.5ex\hbox{\ltsima}}
\def\gtsima{$\; \buildrel > \over \sim \;$}
\def\simgt{\lower.5ex\hbox{\gtsima}}
\def\kms{{\rm\,km \; s^{-1}}}

\def\kpc{{\rm\,kpc}}

\def\msun{{\rm\,M_\odot}}
\def\lsun{{\rm\,L_\odot}}

\def\s{\ifmmode \widetilde \else \~\fi}
\def\={\overline}

\def\spose#1{\hbox to 0pt{#1\hss}}

\def\lta{\mathrel{\spose{\lower 3pt\hbox{$\mathchar"218$}}
     \raise 2.0pt\hbox{$\mathchar"13C$}}}
\def\gta{\mathrel{\spose{\lower 3pt\hbox{$\mathchar"218$}}
     \raise 2.0pt\hbox{$\mathchar"13E$}}}
\def\Dt{\spose{\raise 1.5ex\hbox{\hskip3pt$\mathchar"201$}}}    
\def\dt{\spose{\raise 1.0ex\hbox{\hskip2pt$\mathchar"201$}}}    

\def\dotsfill{\leaders\hbox to 1em{\hss.\hss}\hfill}

\def\Gyr{{\rm\,Gyr}}
\def\FeH{{\rm[Fe/H]}}

\loadboldmathitalic 
\title[The Pristine Dwarf-Galaxy survey II]{The Pristine Dwarf-Galaxy survey -- II. In-depth observational study of the faint Milky Way satellite Sagittarius II.}
\author[N. Longeard et al.] {Nicolas Longeard$^{1}$, Nicolas Martin$^{1,2}$, Else Starkenburg$^{3}$, Rodrigo A. Ibata$^{1}$, 
\newauthor Michelle L. M. Collins$^{4,6}$, Benjamin P. M. Laevens$^{5}$, Dougal Mackey$^{7}$, R. Michael Rich $^{8}$ ,
\newauthor  David S. Aguado$^{9}$, Anke Arentsen$^{3}$, Pascale Jablonka$^{10,11}$, 
\newauthor Jonay I. Gonz\'alez Hern\'andez$^{12,13}$, Julio F. Navarro$^{14}$, Rub\'en S\'anchez-Janssen$^{15,16}$  \\
$^{1}$ Universit\'e de Strasbourg, CNRS, Observatoire astronomique de Strasbourg, UMR 7550, F-67000 Strasbourg, France\\
$^{2}$ Max-Planck-Institut f\"ur Astronomy, K\"onigstuhl 17, D-69117, Heidelberg, Germany\\
$^{3}$ Leibniz Institute for Astrophysics Potsdam (AIP), An der Sternwarte 16, 14482 Potsdam, Germany\\
$^{4}$ Department of Physics, University of Surrey, Guildford, GU2 7XH, Surrey, UK\\
$^{5}$ Institute of Astrophysics, Pontificia Universidad Cat\'olica de Chile, Av. Vicuña Mackenna 4860, 7820436 Macul, Santiago, Chile\\
$^{6}$ Department of Astronomy, Yale University, New Haven, CT 06520, USA\\
$^{7}$ Research School of Astronomy and Astrophysics, Australian National University, Canberra, ACT 2611, Australia\\
$^{8}$ University of California Los Angeles, Department of Physics \& Astronomy, Los Angeles, CA, USA\\
$^{9}$ Institute of Astronomy, University of Cambridge, Madingley Road, Cambridge CB3 0HA, UK\\
$^{10}$ GEPI, Observatoire de Paris, PSL Research University, CNRS, Place Jules Janssen, 92190 Meudon, France\\
$^{11}$ Laboratoire d'astrophysique, \'Ecole Polytechnique F\'ed\'erale de Lausanne (EPFL), Observatoire, 1290 Versoix, Switzerland\\
$^{12}$ Instituto de Astrofisica de Canarias, Via Lactea, 38205 La Laguna, Tenerife, Spain\\
$^{13}$ Universidad de La Laguna, Departamento de Astrofisica, 38206 La Laguna, Tenerife, Spain\\
$^{14}$ Dept. of Physics and Astronomy, University of Victoria, P.O. Box 3055, STN CSC, Victoria BC V8W 3P6, Canada\\
$^{15}$ NRC Herzberg Astronomy and Astrophysics, 5071 West Saanich Road, Victoria, BC V9E 2E7, Canada\\
$^{16}$ Royal Observatory Edinburgh, Blackford Hill, Edinburgh, EH9 3HJ, UK\\
}

\date{\today}
\begin{document} 
\maketitle 
\begin{abstract} 
We present an extensive study of the Sagittarius II (Sgr~II) stellar system using MegaCam $g$ and $i$ photometry, narrow-band, metallicity-sensitive Calcium H\&K doublet photometry, augmented with Keck II/DEIMOS multi-object spectroscopy. We are able to derive and refine the Sgr~II structural and stellar properties: the colour-magnitude diagram implies Sgr~II is old (12.0 $\pm$ 0.5) Gyr and metal-poor. The CaHK photometry confirms the metal-poor nature of the satellite ([Fe/H] $_\mathrm{CaHK} = -2.32 \pm 0.04$ dex) and suggests that Sgr~II hosts more than one single stellar population ($\sigma_\mathrm{[FeH]}^\mathrm{CaHK} = 0.11^{+0.05}_{-0.03}$  dex). From the deep spectroscopic data, the velocity dispersion of the system is found to be $\sigma_{vr} = 2.7^{+1.3}_{-1.0} \kms$ after excluding two potential binary stars. Using the Ca infrared triplet  measured from our highest signal-to-noise spectra, we are able to confirm the metallicity and dispersion inferred from the Pristine photometric metallicities:  ([Fe/H]$_\mathrm{spectro} = -2.23 \pm 0.05$ dex,  $\sigma_\mathrm{[Fe/H]}^\mathrm{spectro} =  0.10 ^{+0.06}_{-0.04} $ dex). Sgr~II's metallicity and absolute magnitude (M$_V = -5.7 \pm 0.1$ mag) place the system on the luminosity-metallicity relation of the Milky Way dwarf galaxies despite its small size. The low, but resolved metallicity and velocity dispersions paint the picture of a slightly dark matter-dominated satellite. Furthermore, using the \textit{Gaia} Data Release 2, we constrain the orbit of the satellite and find an apocenter of $118.4 ^{+28.4}_{-23.7} \kpc$ and a pericenter of $54.8 ^{+3.3}_{-6.1} \kpc$. The orbit of Sgr~II is consistent with the trailing arm of the Sgr stream and indicate that it is possibly a satellite of the Sgr dSph that was tidally stripped from the dwarf's influence.

\end{abstract}
 

\section{Introduction}

During the history of the Universe, structures such as galaxies formed hierarchically. Therefore, dwarf galaxies (DGs) are particularly old and metal-poor systems and targets of choice to study the history of the local universe. They are systems spanning a wide range of masses and luminosity. Bright dwarf galaxies such as Sculptor \citep{shapley38b}, Draco \citep{wilson55}, or Sextans \citep{irwin90} have been known for decades \citep{mateo98}, but the extensive search for these systems over the last twenty years revealed fainter and fainter systems (\citealt{martin06}, \citealt{belokurov07}, \citealt{zucker06b}). Still, our knowledge of the Milky Way satellites remains incomplete. The recent discoveries of several of those faint galaxy candidates with $M_V > -4$ \citep{willman05a, belokurov07, drlica-wagner15, laevens15b, luque16} are promising as they might well bring new perspectives to near-field cosmology \citep{bullock17}.

The study of these nearby small-scale structures can allow one to explore various problematics in astrophysics, from the faint-end of the galaxy luminosity function \citep{koposov09} to the validity of cosmological models.  Therefore, DGs are important cosmological probes (\citealt{pawlowski17,tulin17}) as the comparison of their observed properties with the predictions made by our current $\Lambda$CDM model leads to some discrepancies that we have to understand in order to constrain and refine our cosmological models. DGs are also thought to be among the most dark matter (DM hereafter) dominated systems in the universe \citep{wolf10} and could be useful for the detection of the elusive DM particle through self-annihilation processes (\citealt{bertone05, geringer-sameth15a}). 
However, using faint dwarf galaxies as cosmological probes can be challenging as their exceptional faintness comes with observational challenges. The overall properties and/or even the very nature of the recently discovered systems can sometimes be puzzling as the distinction between galaxy and globular cluster is difficult to make (\citealt{conn18}, \citealt{longeard18}). Therefore, only the combined efforts of deep photometric surveys, such as the Dark Energy Survey \citep[DES]{abbott05}, the Panoramic Survey Telescope and Rapid Response System \citep[PS1]{chambers16}, or the Sloan Digital Sky Survey \citep[SDSS]{york2000}, and spectroscopic observations can hope to improve our understanding on the faint-end of the luminosity function and the history of the Milky Way (MW).  \\

In this context, we present here the study of the Milky Way satellite Sgr~II, discovered by \citet[ hereafter L15]{laevens15} in PS1, where it was identified as an old (12.5 Gyr) and metal-poor ([Fe/H] = -2.20 dex) dwarf-galaxy candidate. L15 noticed that Sagittarius II had a peculiar location on the sky: its position and distance were found to be consistent with the predictions of models for the Sagittarius stream \citep{law_majewski10}. They concluded that this satellite might actually have been a satellite of the bright Sagittarius dwarf galaxy discovered by \citet{ibata94}. However, spectroscopic observations were still needed at the time to dynamically tie the stream and Sagittarius II, as well as confirming the galaxy nature of the satellite. Sgr~II was also studied by \citet[M18]{mutlu_pakdil18} with Magellan/MegaCam photometry, who confirmed the structural properties inferred from L15. Furthermore, using both blue horizontal branch stars (BHBs) and a CMD-fitting technique, they found an old ($13.5$ Gyr), metal-poor ($\FeH = -2.2$ dex) stellar population, with an alpha abundance ratio of  $\alpha$/Fe $= 0.4$ dex, and a distance modulus $m-M = 19.2 \pm 0.2$ mag. Moreover, they found a half-light radius of $32 \pm 1.0$ pc, and an absolute magnitude of $M_V = -5.2 \pm 0.1$ mag. Based on all these properties, M18 concluded that the system was likely a globular cluster, and compared the satellite to several extended clusters of M31 associated to known streams, in the same way that Sgr~II is suspected to belong to the Sgr stream. However, M18 emphasised the importance of a spectroscopic study to confirm their conclusion.

In this work, we show a thorough analysis of the stellar, structural and orbital properties of Sagittarius II, using deep broadband photometry from the Canadian-France-Hawaii Telescope (CFHT) MegaCam (MC hereafter) imager in the context of Pristine. The Pristine survey uses a narrow-band filter centred on the metallicity-sensitive Ca H\&K doublet (\citealt{starkenburg17}) to identify metal-poor stars and estimate their metallicity using pure photometry. Keck II/DEIMOS spectroscopy are additionally used to constrain the system's metallicity and kinematics. Finally, combined with the \textit{Gaia} Data Release 2, we constrain the orbital properties of the satellite.

\section{Observations and Data}
\subsection{Photometry}

Our photometry consists of deep broadband $g_\mathrm{MC}$ and $i_\mathrm{MC}$ observations as well as narrow-band observations with the CaHK filter centred on the metallicity-sensitive Calcium H$\&$K doublet. It was observed using the wide-field imager MegaCam on the CFHT \citep{boulade03}. The CaHK photometry is part of a larger survey called Pristine \citep{starkenburg17}. 
The ongoing Pristine and Pristine dwarf-galaxy surveys aim to probe the northern part of the MW halo, the MW bulge, as well as many of the northern faint dwarf galaxies and dwarf-galaxy candidates.  A study of the MW satellite Draco II, from a very similar data set to that of the current paper, is presented in \citet[ hereafter L18]{longeard18}.

Observations were conducted in service mode by the CFHT staff during the night of July, 2nd, 2016 under good seeing conditions ($\sim 0.4''$). We refer to reader to L18 for the details of the Megacam data reduction. The star/galaxy separation is done using the Cambridge Astronomical Survey Unit \citep{irwin01} pipeline flags, which also indicate saturated sources. The MegaCam photometry is calibrated onto the PS1 photometric system with the same procedure as in L18: a cross-identification of all unsaturated point sources with photometric uncertainties below 0.05 mag in both catalogs is performed. The difference $g_\mathrm{MC} - g_\mathrm{PS1}$ (respectively $i_\mathrm{MC} - i_\mathrm{PS1}$) is expressed as a function of the colour $g_\mathrm{MC} - i_\mathrm{MC}$. We then fit a third-order polynomial to translate MC photometry into PS1 through a 3$\sigma$ clipping procedure. The coefficients of the polynomials to transform ($g_\mathrm{MC}$,$i_\mathrm{MC}$) into ($g_\mathrm{PS1}$,$i_\mathrm{PS1}$) in this work are different from those in L18.  We define $x \equiv g_{MC} - i_{MC}$ and obtain:

\begin{eqnarray}
g_\mathrm{MC} - g_\mathrm{P1} &  = & a^g_0x^{2}  + a^g_1x + a^g_2 ,\nonumber\\
i_\mathrm{MC} - i_\mathrm{P1} & = & a^i_0x^{2}  + a^i_1x + a^i_2,\nonumber
\end{eqnarray}

\noindent with $x \equiv g_\mathrm{MC} - i_\mathrm{MC}$. The calibration coefficients are: $a^g_0 = -0.0162 \pm 0.0046$, $a^g_1 = 0.0906 \pm 0.0029$, $a^g_2 = -0.0696 \pm 0.0016$ for the $g$ band and $a^i_0 = -0.0117 \pm 0.0032$, $a^i_1 = 0.0058 \pm 0.0022$, $a^i_2 = -0.1359 \pm 0.0010$ for the $i$ band. All uncertainties on the polynomials coefficients are propagated into the photometric uncertainties. 

All stars saturated in the MC photometry, filtered during the calibration process, are taken directly from PS1 and added to the final catalog, for a total of 83,355 stars. This catalog is finally dereddened using the dust map from \citet{schlegel98} and the extinction coefficients from \citet{schlafly11}. In the rest of the text, we use the combined catalogue and the PS or MC subscripts are dropped.

\subsection{Spectroscopy}

\begin{figure}
\begin{center}
\centerline{\includegraphics[width=\hsize]{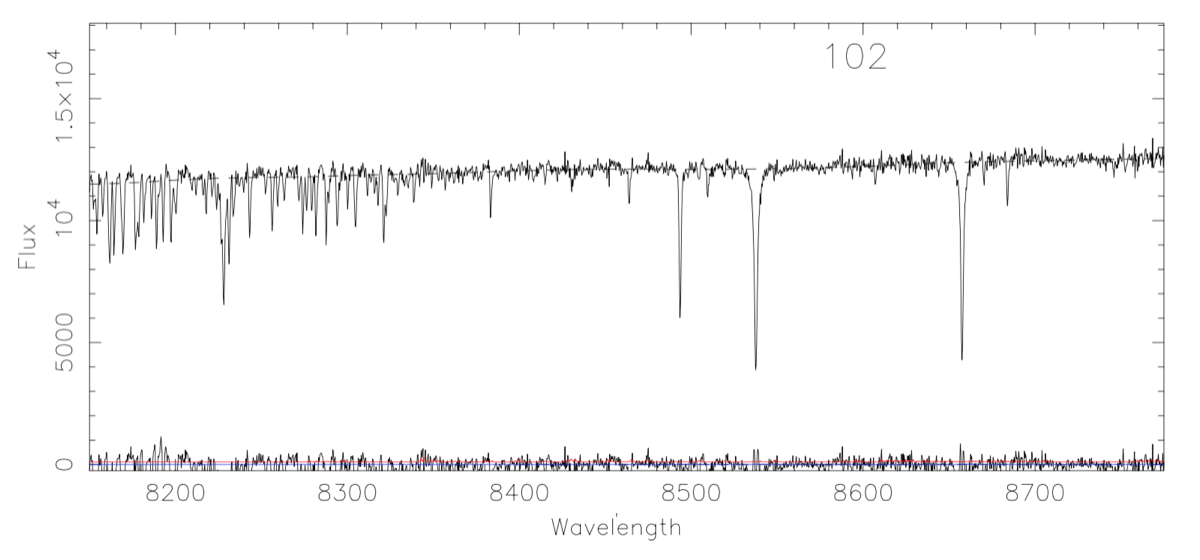}}
\caption{Spectrum of a $g_0 \sim 17.5$ star in our spectroscopic data set, corresponding to a S/N $\sim 90$. This spectrum is already processed and sky-substracted.}
\label{spectrum} 
\end{center}
\end{figure}

Spectroscopic follow-up observations of Sgr~II were obtained with Keck/DEIMOS \citep{faber03}: ``mask 1'' was observed on the 2015-09-18, ``mask 2'' on the 2015-09-08 and ``mask 3'' on the 2015-09-12 . Mask 3 is a re-observation of mask 1, while mask 2 was chosen perpendicular to the other two in order to probe potential Sgr~II members further away in the South/North direction (Figure \ref{field}).  We refer the reader to \citet{ibata11} for a detailed description of our DEIMOS data reduction procedure. All stars with a signal-to-noise ratio (S/N) below 3 or with a velocity uncertainty greater than 15 km s$^{-1}$ were discarded. Following the procedure described in \citep{simon_geha07} and using the 51 stars observed at least twice, we assess the systematics in our sample and find a negligible bias of $0.4 \pm 1.3 \kms$ and a systematic uncertainty floor of $\delta_{thr} = 1.8 ^{+0.3}_{-0.2} \kms$.  The heliocentric velocities of each star observed more than once are merged into one single measurement  by taking the mean of all available quantities weighted by the inverse of their respective velocity uncertainties. The same procedure was followed for the equivalent widths of the Ca triplet. For illustrative purposes, an example of spectrum of S/N $\sim$ 90 is shown in Figure \ref{spectrum}.

Finally, the existence of binaries in the sample is investigated for all stars with multiple velocity measurements. To do so, we define the quantity $\mu$ such that

\begin{eqnarray}
\mu = \frac{v_{r,l} - v_{r,m}}{\sqrt{\delta_{vr,l}^2 + \delta_{vr,m}^2 + 2\delta_{thr}^{2}}}, \nonumber
\end{eqnarray}

\noindent with $v_{r,l}$ the heliocentric velocity of a star in the mask l (resp. for $v_{r,m}$), and $\delta_{vr,l}$ the uncertainty on this measurement (resp. for $\delta_{vr,m}$). If $\mu$ is greater than 2.5, the star is considered as a possible binary and flagged accordingly. Two stars, observed in mask 1 and 3, are identified as binaries trough this procedure, with differences in velocities of $21.46 \kms$ and $25.07 \kms$.

\section{Broadband photometry analysis}
\begin{figure*}
\begin{center}
\centerline{\includegraphics[width=\hsize]{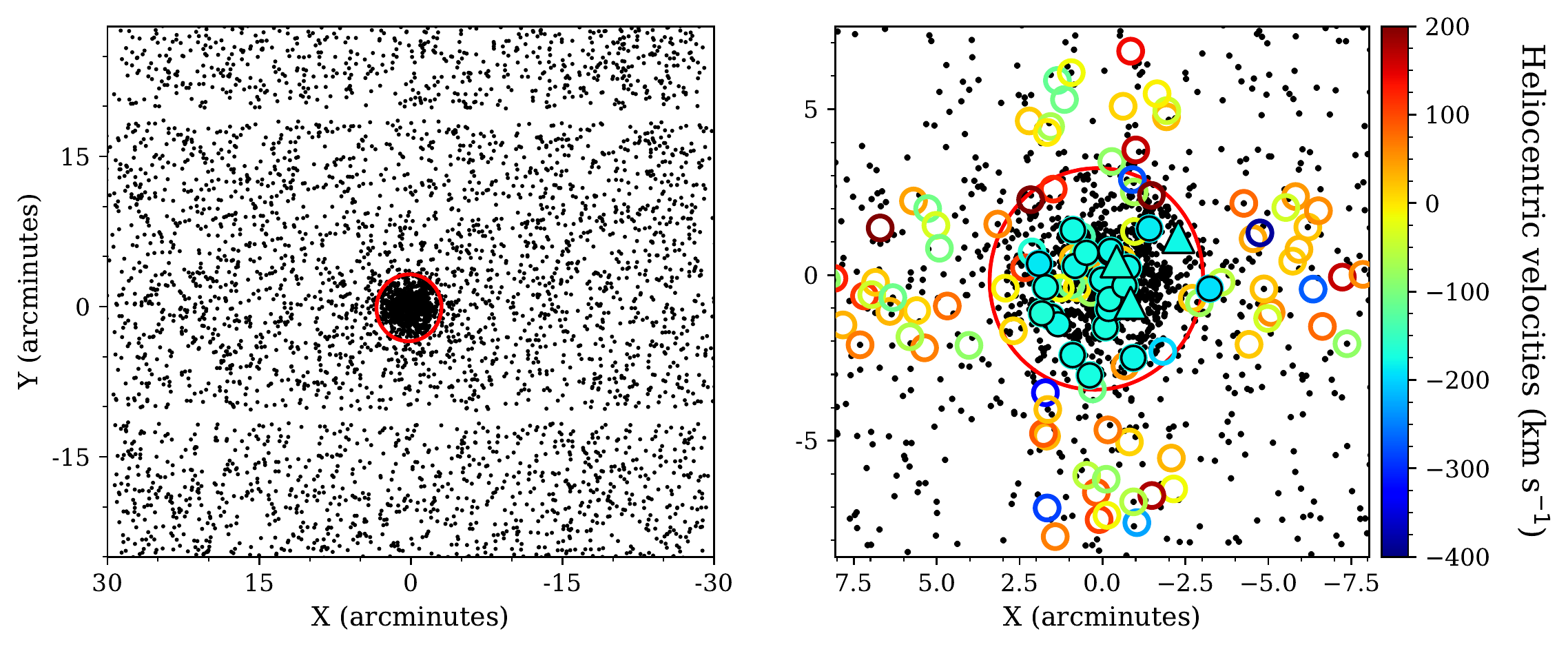}}
\caption{{\textit{Left panel: }}  Spatial distribution of Sgr~II-like stars, centred on ($\alpha_0 = 298.166284027^\circ, \delta_0 = -22.89632528^\circ$). The red contour defines the two half-light radii ($r_h \sim 1.7'$) of the satellite.  {\textit{Right panel: }} Close-up on the central region, with stars observed spectroscopically are colour-coded according to their heliocentric velocities. Filled circles represent very likely Sgr~II members, while filled triangles represent Sgr~II HB stars in the spectroscopic data set.}
\label{field} 
\end{center}
\end{figure*}

We present the one square degree field centred on Sgr~II together with the spatial distribution of stars observed with spectroscopy in Figure \ref{field}. 

\begin{figure*}
\begin{center}
\centerline{\includegraphics[width=\hsize]{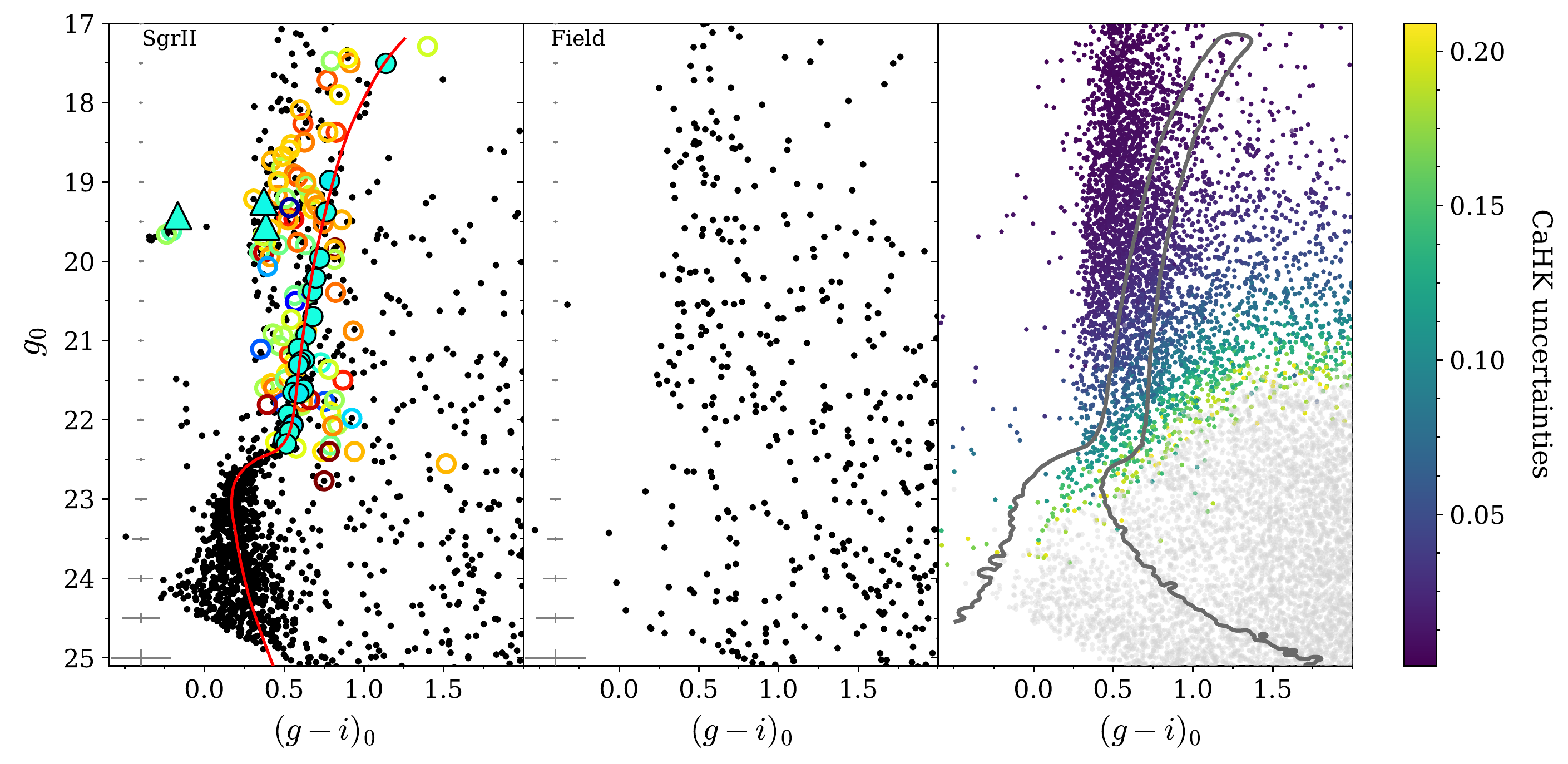}}
\caption{ {\textit{Left panel:}}  CMD within two $r_\mathrm{h}$ of Sgr~II. Its old ($>$ 10 Gyr) and metal-poor ([Fe/H] $<$ -2.2) stellar population clearly stands out. One can notice the presence of a few blue stragglers in the system around $g_0 \sim 22.0$, as well as Sgr~II horizontal branch. Stars observed with spectroscopy are shown with coloured circles. The filled ones represent the confirmed spectroscopic members. Filled triangles show the location, in the CMD, of HB stars in the spectroscopic data set. The favoured isochrone for Sgr~II, obtained in section 3, is shown as a solid red line ($12 \Gyr$, $\FeH = -2.35$ dex, $\alpha$/Fe $= 0$ dex, $m-M$ $= 19.32$ mag). {\textit{Middle panel:}} CMD of the field for an equivalent area. {\textit{Right panel:}} Photometric uncertainties for the $CaHK$ band. The grey contours show the mask used to select the Sgr~II-like population showed in the spatial distribution of Figure \ref{field}.}
\label{CMDs} 
\end{center}
\end{figure*}

The colour-magnitude diagram (CMD) of all stars within two half-light radii ($r_h \sim 1.7'$) of the system is shown in the left panel of Figure \ref{CMDs}, along with the spectroscopically observed stars. The CMD of the same areal coverage but selected in the outskirts of the MegaCam field of view is represented as a comparison in the middle panel. Sgr~II's main sequence (MS) and main sequence turn-off (MSTO) are very well defined thanks to the depth of the MegaCam data, and corresponds to an old ($>$ 12 Gyr) and very metal-poor ([Fe/H] $< -2.0$) population. A few blue stragglers can be seen in the satellite. Sgr~II also hosts a few blue horizontal branch stars around $g_0 \sim 19.7$. 

The BHB stars are useful as they are good distance tracers \citep[D11]{deason11}. First, all 12 BHBs' $g_0$ and $r_0$ are calibrated onto the SDSS photometry according to the colour equations of \citet{tonry12}. The median absolute magnitude of the BHBs is obtained using equation (7) of D11 and yields $M_g = 0.47 \pm 0.02$ mag, yielding a median distance modulus of $(m-M)_\mathrm{BHB} = 19.19 \pm 0.10$ mag or $68.8 \pm 3.0$ kpc.

\subsection{Structural and CMD fitting}
We use our MC photometry to refine the structural properties of Sgr~II previously studied by L15 and M18 and determine its main stellar properties through a CMD and spatial distribution fitting procedure. The formalism of this analysis is detailed in \citet{martin16} and L18. Though the main steps will be briefly detailed below, we refer the reader to these two references for more details. Six structural parameters will be inferred from our analysis: the centroid offsets along the X and Y axes with respect to the centre coordinates of the literature, $x_0$ and $y_0$, in arcminutes,  the ellipticity $\epsilon$ \footnote{The ellipticity is defined as 1 - $\frac{a}{b}$, with $a$ and $b$ the major and minor axis. }, the half-light radius $r_h$, the position angle $\theta$, and the number of stars $N^*$ of the satellite. These structural properties are gathered in a parameter set noted $\mathcal{P}_\mathrm{spac} \equiv \{x_{0},y_{0}, \epsilon, r_{h},  \theta, N^*\}$.

We then define the CMD parameters derived by our CMD fitting procedure: the age of the satellite $A$, the systemic metallicity $\FeH_\mathrm{CMD}$, the [$\alpha$/Fe] abundance ratio, the distance modulus $m-M$, and $\eta$ the fraction of Sgr~II stars with respect to the total number of stars in the field chosen for the anlaysis.  We regroup these properties into the set $\mathcal{P}_\mathrm{CMD} \equiv \{ A, \FeH_{CMD}, [\alpha/Fe],m-M,\eta  \}$

For a given star $k$, we consider its following properties: its $g_k$ and $i_k$ magnitudes, and its position offset from the center coordinates of the literature, $X_k$ and $Y_k$. These four properties are gathered into one set $\vec{d_k} = \{X_k,Y_k,g_k,i_k\}$. \\

The Sgr~II radial density, $\rho_\mathrm{dwarf}$, is modelled by an exponential radial profile while the foreground contamination is assumed constant over the field of view, i.e.

\begin{equation}
    \rho_\mathrm{dwarf}(r) = \frac{1.68^{2}}{2 \pi r_h^2(1-\epsilon)}\exp(-\frac{1.68r}{r_h}),
\end{equation} 

\noindent with $r$ the elliptical radius, which can be expressed using the projected sky coordinates $(x,y)$ as

\begin{eqnarray}
\label{eqn:r}
r=\bigg(  \Big(\frac{1}{1-\epsilon} \Big((x-x_0)\cos\theta - (y-y_0)\sin\theta\Big)\Big)^2\nonumber\\
+ \Big((x-x_0)\sin\theta +(y-y_0)\cos\theta\Big)^2  \bigg)^{1/2}.
\end{eqnarray}

For the $k$-th star, the spatial likelihood can then simply be written as

\begin{equation}
\ell_\mathrm{sp}^\mathrm{SgrII}(X_k,Y_k)  =  \frac{\rho_\mathrm{dwarf}(r)}{\int_{S} \rho_\mathrm{dwarf}(r) dS },
\end{equation}

\noindent where $S$ in the area of the sky over which the analysis is conducted.

The spatial likelihood of the Milky Way foreground contamination is assumed flat, which yields

\begin{equation}
\ell_\mathrm{sp}^\mathrm{MW}  =   \frac{1}{\int{dS}}.
\end{equation}

The CMD likelihood function $\ell_\mathrm{CMD}$ is built from the sum of two models: one for the foreground, $\ell_\mathrm{CMD}^\mathrm{MW}$, constructed empirically from the field CMD, and one to describe the Sgr~II population taken as a single star population, and called $\ell_\mathrm{CMD}^\mathrm{SgrII}$. The foreground contamination model is built by selecting all stars outside 5$r_h$ of the system centre and binning their distribution on the CMD. This distribution is smoothed by a gaussian kernel in both colour and magnitude of a width of 0.1 in an attempt to limit the effects of shot noise.
$\ell_\mathrm{CMD}^\mathrm{SgrII}$ is generated using a range of Darmouth isochrones \citep{dotter08}. We choose isochrones of different $\FeH_\mathrm{CMD}$, $A$, [$\alpha$/Fe], and distance modulus $m-M$. The priors on each parameters are reported in Table 1. To build the PDF of a given stellar population, we simulate the CMD of a population of several million stars, based on its isochrone, luminosity function and photometric uncertainty at a given ($g_0$,$i_0$). To avoid aliasing effects, especially at the bright end of our models where the photometric uncertainties are unrealistically low, we add 0.01 in quadrature to the photometric uncertainties. Finally, each PDF is degraded to the completeness of the data estimated by following the method of \citet{martin16}. The 50 per cent completeness is reached at $g_0 \sim 24.2$ and $i_0 \sim 23.4$ mag.

The structural and CMD parameters are gathered into a single set $\mathcal{P} \equiv \{A, [Fe/H], [\alpha/Fe], \mu, \epsilon, r_{h}, x_{0},y_{0}, \theta,\eta\}$. At the distance of Sgr~II, the tip of the Red Giant Branch (RGB) tip is expected to be at $g_0 \sim 17.0$. Furthermore, misidentified background galaxies start to pollute our photometry below $g_0 \sim 25.0$. The fit does not take into account the horizontal branch stars as these are poorly modelled by the theoretical stellar population models. Therefore, the analysis is restricted in a specific CMD box defined with the following cuts: $-0.2 < g_0 - i_0 < 1.2$ and $17.0 < g_0 < 25.0$. CMD and spatial properties are fitted at the same time through our own Markov Chain Monte Carlo \citep[MCMC]{hastings70} algorithm by maximising the likelihood of the following model:

\begin{equation}
\mathcal{L}_{tot} = \sum^{N}_{k=1}  \ell_{tot}(\vec{d_{k}} | \mathcal{P}) = \eta \ell_\mathrm{SgrII}(\vec{d_{k}} | \mathcal{P}) + (1 - \eta ) \ell_\mathrm{MW}(\vec{d_{k}}),
\end{equation} 

\noindent with

\begin{eqnarray}
\ell_\mathrm{SgrII}(\vec{d_k} | \mathcal{P}) & = & \ell_\mathrm{CMD}^\mathrm{SgrII}(g_{k},i_{k} | \mathcal{P}_\mathrm{CMD}) \ell_\mathrm{sp}^\mathrm{SgrII}(X_{k},Y_{k} | \mathcal{P}_\mathrm{spac}), \\
\ell_\mathrm{MW} & = & \ell_\mathrm{CMD}^\mathrm{MW}(g_{k},i_{k}) \ell_\mathrm{sp}^\mathrm{MW}(X_{k},Y_{k}).
\end{eqnarray} 

\begin{figure*}
\begin{center}
\centerline{\includegraphics[width=\hsize]{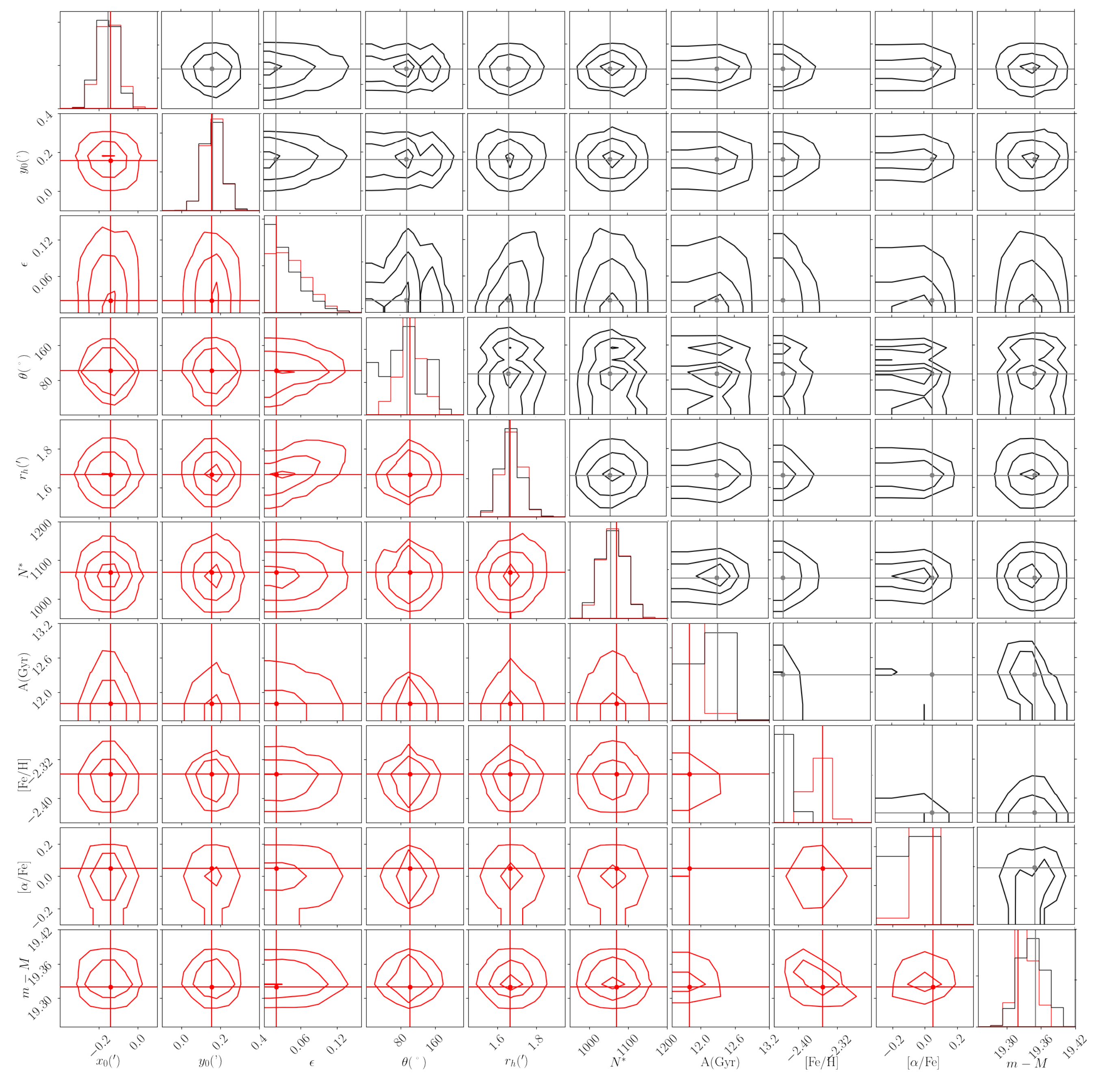}}
\caption{One and two-dimensional PDFs of the structural and CMD properties of Sgr~II, inferred using the method described in section 3.1. Red lines correspond to the favoured inference, using both the distance prior based on BHBs and the metallicity prior from the spectroscopy, while the black contours show the case without any prior. These contours are defined as the usual 1, 2 and 3$\sigma$ confidence intervals in the 2D, gaussian case.}
\label{cornerplot} 
\end{center}
\end{figure*}

Finally, the distance to Sgr~II is constrained using a Gaussian prior based on the distance modulus derived from the median absolute magnitude of the BHBs in the first paragraph of section 3 ($m-M = 19.19 \pm 0.10 $ mag). A Gaussian prior on the metallicity of the satellite is also applied and comes directly from the combination of the spectroscopic and CaHK metallicity measurements detailed in the sections 4 and 5 respectively ($\FeH_\mathrm{SgrII} = -2.28 \pm 0.03$ dex).
The inference of each parameter of $\mathcal{P}$ is summed up in Table 1, and the 2D PDFs  are shown in Figure \ref{cornerplot}.

\begin{table*}
\renewcommand{\arraystretch}{1.1}
\begin{center}
\caption{Properties of Sgr~II.\label{tbl-2}}
\begin{tabular}{lcccc}
\hline
Parameter &  Unit & Prior & Favoured model & Uncertainties  \\
\hline
Right ascension (ICRS) $ \alpha $ & degrees & --- & 298.166284027 & $\pm 0.0008$ \\
Declination (ICRS) $ \delta $ & degrees & --- & $-22.89632528$  & $\pm 0.0008$ \\
$l$ & degrees & --- & 18.93203283 & $\pm 0.0008$ \\
$b$ & degrees & --- & $-22.89360512$  & $\pm 0.0008$ \\
$r_{h}$ & arcmin & $> 0$ & 1.7 & $ \pm 0.05 $  \\
$r_{h}$ & pc & & 35.5 & $^{+1.4}_{-1.2}$ \\
$\theta$ & degrees & [0,180] & 103 & $^{+28}_{-17}$ \\
$\epsilon$ & --- & $[0,1]$ & 0.0 & $< 0.12$ at the $95$\% CL  \\
Distance modulus & mag & [18.90,19.45] & 19.32 & $^{+0.03}_{-0.02}$  \\
Distance & kpc & & 73.1 & $^{+1.1}_{-0.7}$  \\
$A$ & Gyr & [9,13.5] & 12.0 & $\pm$ 0.5  \\
$\FeH$ & dex & [-2.4,-1.5] & $-2.28$ & $\pm$ 0.03  \\
$[\alpha/\textrm{Fe}]$ & dex & [-0.2,0.6] & 0.0 & $\pm 0.2$ \\
log(Luminosity) & --- & $> 0$ & $4.20$ & $\pm 0.1$  \\
$M_V$ & mag & --- & -5.7 &  $\pm$ 0.1 \\
$\mu_{0}$ & mag arcsec$^{-2}$ & --- & $24.7$ & $\pm 0.15$  \\
$<v_r>$ & $\kms$ & --- & $-177.3$ & $ \pm 1.2$  \\
$\sigma_{vr}$ & $\kms$ & $>$ 0 & $2.7$ & $^{+1.3}_{-1.0}$  \\
$\mu_{\alpha}^{*}$ & mas yr$^{-1}$  & --- & $-0.65$ & $^{+0.08}_{-0.10}$   \\
$\mu_{\delta}$ & mas yr$^{-1}$ &  --- & $-0.88$ & $\pm 0.12$   \\
Apocenter & kpc  & --- & $118.4$ & $^{+28.4}_{-23.7}$   \\
Pericenter & kpc &  --- & $54.8$ & $^{+3.3}_{-6.1}$   \\
$\epsilon_{\mathrm{orbit}}$& ---  & $> 0$ & 0.44 & $\pm 0.01$  \\
$U$ & $\kms$ & ---  & 0.4  & $^{+14.1}_{-19.5}$   \\
$V$ & $\kms$ & ---  & $-366.5$ & $^{+27.3}_{-42.8}$ \\
$W$ & $\kms$ & ---  & 160.3 & $^{+26.3}_{-19.9}$  \\
$L_z$ & km s$^{-1}$ kpc  & ---  & 6292 & $^{+2236}_{-1899}$ \\
$E$ & km$^{2}$ s$^{-2}$ & ---  & 17159 & $^{+10213}_{-3120}$  \\
\hline
\end{tabular}
\end{center}
\end{table*}

\begin{figure}
\begin{center}
\centerline{\includegraphics[width=\hsize]{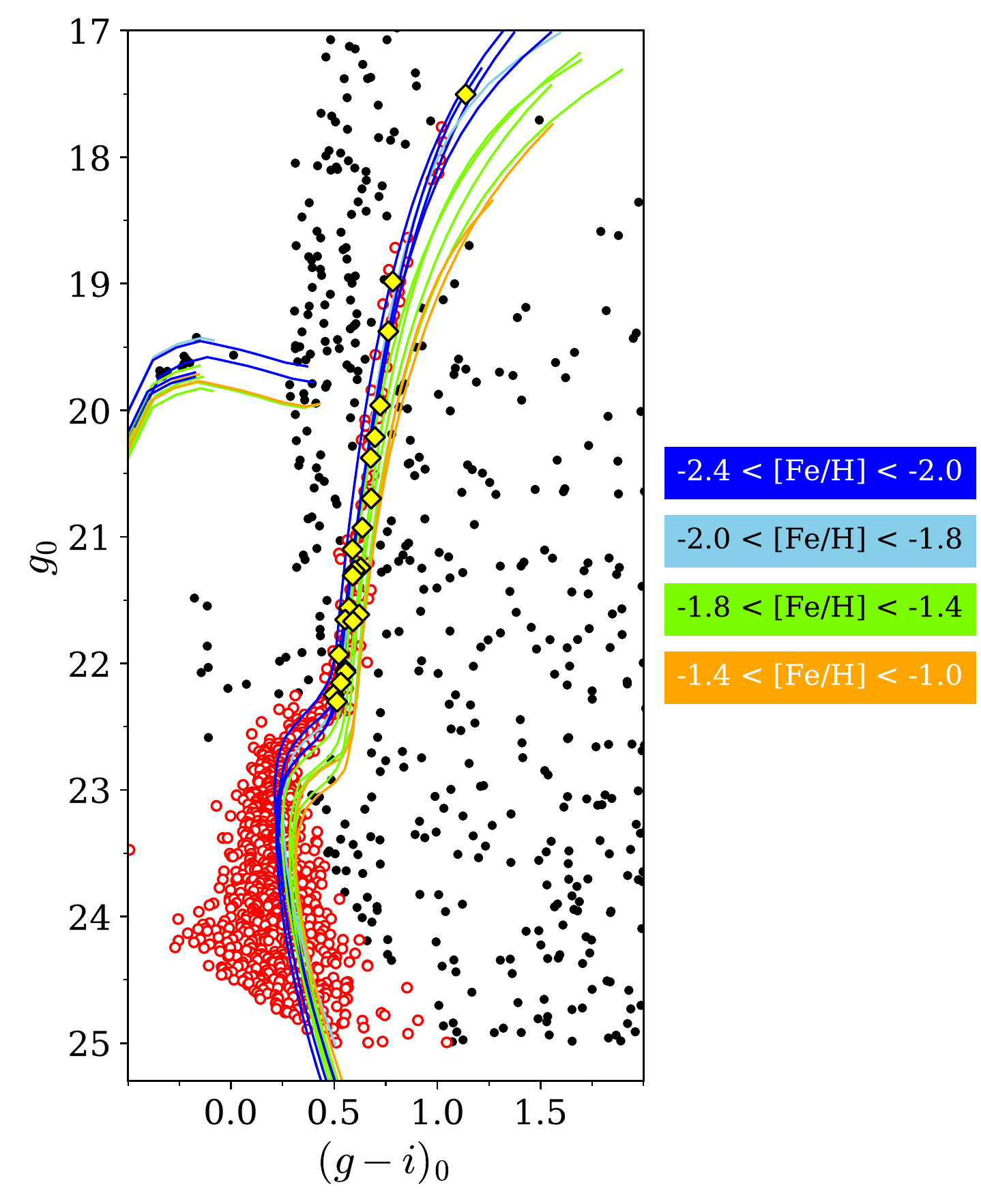}}
\caption{CMD within two half-light radii of Sgr~II overplotted with the fiducials of MW globular clusters from \citet{bernard14}. Stars with a Sgr~II membership probability greater than ten per cent are shown as red circles. Spectroscopically confirmed members of Sgr~II are shown as yellow diamonds. The fiducials are separated in metallicity bins, from the most metal-poor in dark blue to the most metal-rich available in orange, and shifted to the favoured distance modulus $19.32$ found for Sgr~II. }
\label{fiducials} 
\end{center}
\end{figure}

The best-fit isochrone is shown as the red PDFs in Figure \ref{cornerplot}. Sgr~II is found to be significantly old and metal-poor with an age of $12.0 \pm 0.5$ Gyr population along with a systemic metallicity of $-2.35 \pm 0.05$ dex. Furthermore, the alpha abundance of this isochrone is solar ([$\alpha$/Fe] = 0.0), though we caution the reader about reading too much into this parameter given the roughness of the [$\alpha$/Fe] abundance ratio grid. Finally, the favoured distance modulus is $\mu =  19.32^{+0.03}_{-0.02}$ mag, and corresponds to a physical distance of $73.1^{+1.1}_{-0.7}$ kpc. We compare these results by performing the fit without the BHBs or the spectroscopic metallicity priors. For this case, the PDFs are shown in black in Figure \ref{cornerplot}. The inferences of all the parameters are compatible: the stellar population is here found to be older ($12.5 \pm 0.5 \Gyr$), more metal-poor ($\FeH ~ -2.45$, at the lower edge of the metallicity grid) and with a distance modulus of $m-M$ = $19.35 \pm 0.05$ mag.  All structural properties are perfectly compatible with L15 and M18: Sgr~II is consistent with being spherical ($\epsilon < 0.12 $ at the 95\% CL) and has a size of $r_h = 1.70 \pm 0.05$ arcminutes, translating into a physical size of 35.5 $^{+1.4}_{-1.2}$ pc. All the main properties of Sgr~II are summarised in Table 1.

Finally, we investigate the presence of RR Lyrae in the field by cross-identifying the PS1 RR Lyrae catalog of \citet{sesar17} with our photometry. Three RRLyrae are found in the vicinity of Sgr~II. Two of these have similar distance modulii, as inferred from \citet{sesar17} (18.73 and 18.85 mag). However, the resulting distances are discrepant from both our BHB and CMD fitting analyses by 0.5 mag (roughly 10 kpc in physical distance). To confirm the distance modulus of Sgr~II, we compare the CMD of the satellite with fiducials of MW globular clusters in PS1 \citep{bernard14} in Figure \ref{fiducials}. In this plot, all fiducials are deredenned and their distance modulii are corrected to correspond to our favoured model for Sgr~II ($m-M$ $ = 19.32 \pm 0.03$ mag). The most metal-poor fiducials are undoubtedly the ones that better reproduce the features of the CMD of Sgr~II, including the HB. Shifting the fiducials to the distance of the two RR Lyrae in the field would allow Sgr~II features to be best reproduced by the more metal-rich fiducials in green. However, Sgr~II's metallicity is clearly lower than that as shown by our CMD fitting as well as the spectroscopic and CaHK analyses respectively in section 4 and 5. Therefore, the distance implied by the two RR Lyrae is clearly not that of Sgr~II. A plausible origin for these two stars might just be the Sgr stream, as shown in section 6.

\subsection{Luminosity}
\begin{figure}
\begin{center}
\centerline{\includegraphics[width=\hsize]{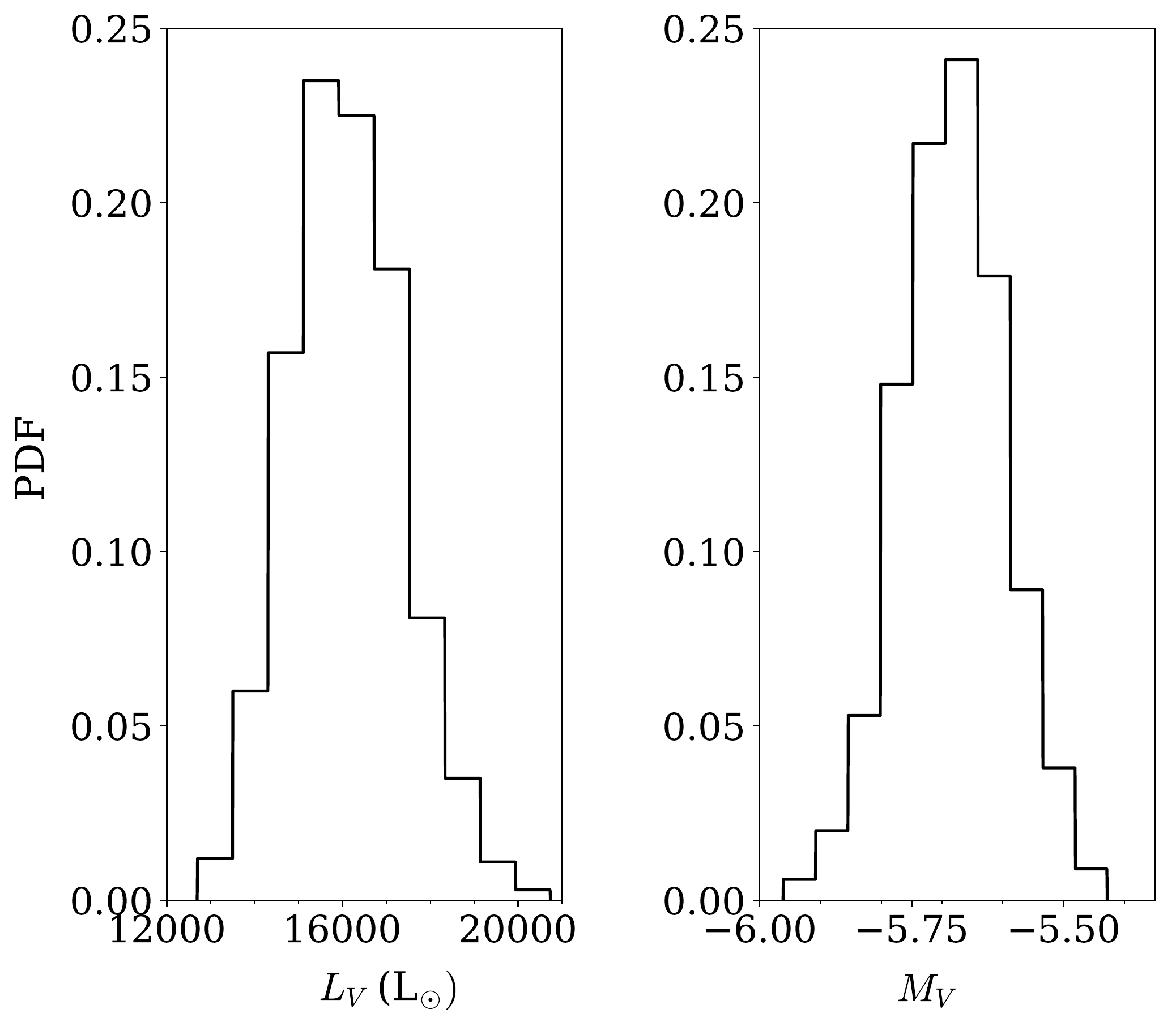}}
\caption{PDFs of the Sgr~II luminosity and absolute magnitude. The favoured luminosity of the satellite is $log(L_V) = 4.2 \pm 0.1$, corresponding to an absolute magnitude of $M_V = -5.7 \pm 0.1$ mag.}
\label{pdfs_lum} 
\end{center}
\end{figure}

The luminosity, absolute magnitude, and surface brightness of Sgr~II are derived using the formalism of \citet{martin16}. The first step consists of drawing a set of parameters denoted $j$ from the final multi-dimensional PDF obtained through the analysis of section 3.1. These parameters are the number of stars N$^*_j$, an age $A_j$, the metallicity $\FeH^{CMD}_{j}$, the alpha abundance ratio [$\alpha/Fe$]$_j$, and the distance modulus $(m-M)_j$. A CMD of the $j$-th stellar population is then simulated; for each simulated star, we ensure that its location in the colour-magnitude diagram falls in the CMD box used to perform the structural and CMD fit ($-0.2 < g_0 - i_0 < 1.2$ and $17.0 < g_0 < 25.0$). Furthermore, a completeness test is performed: the completeness of the survey is first estimated at the colour and magnitude of the simulated star. Then, two random numbers $a$ and $b$ between 0 and 1 are drawn:  if the completenesses of the star in both $g$ and $i$ is greater than these numbers, it is flagged. When the number of flagged stars reaches $N^*_j$, the fluxes of all simulated stars, flagged or not, are summed, which gives the luminosity $L_j$ of the satellite for the $j$-th iteration. This procedure is repeated a thousand times in order to have visually pleasant PDFs. 

The 1D marginalised PDFs of Sgr~II's luminosity and absolute magnitude $M_V$ are represented in Figure \ref{pdfs_lum}. The final favoured parameters are reported in Table 1. The luminosity of the satellite is found to be $\mathrm{log}(L_V) = 4.2 \pm 0.1 $. This measurement is in agreement with both L15 and M18 ( $\mathrm{log}(L_V) = 4.1 \pm 0.1$). Finally, we obtain a surface brightness of $S_0 = 24.7 \pm 0.15 $ mag arcsec$^{-2}$.

\section{Narrow-band CaHK photometry analysis} 

\begin{figure}
\begin{center}
\centerline{\includegraphics[width=\hsize]{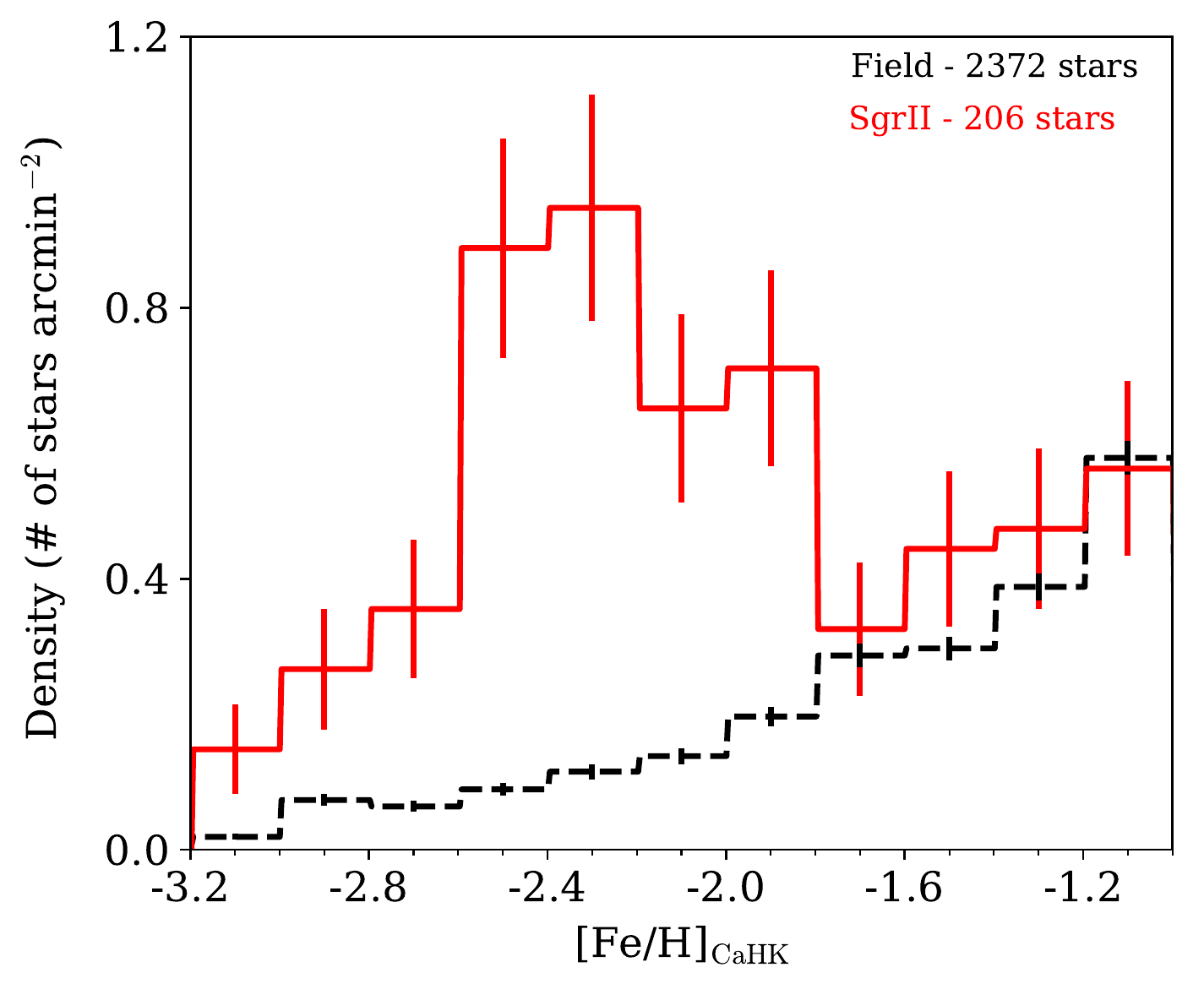}}
\caption{Normalised distribution of Pristine photometric metallicities for all stars within 2r$_h$ (solid red line). The same histogram is also shown for all field stars, i.e. stars outside 5r$_h$ (black dashed line). Sgr~II clearly peaks at [Fe/H]$_\mathrm{CaHK} \sim -2.3$ dex, while no such overdensity exists for the field distribution. }
\label{histos_FeH} 
\end{center}
\end{figure}

The CaHK photometry can be used to estimate the metallicity of Sgr~II and its metallicity dispersion. The photometric metallicity of each star from ($CaHK_0$,$g_0$,$i_0$) can be estimated using the model detailed in \citet{starkenburg17}. Pristine observations are shallower than our broadband $g$ and $i$ photometry (right panel of Figure \ref{CMDs}) and therefore can only be used to estimate the photometric metallicity [Fe/H]$_\mathrm{CaHK}$ of stars in our field down to $g_0 \sim 23$ mag. \citet{starkenburg17} show that the Pristine metallicities are slightly biased low as we go toward the metal-poor end of the calibration model. Therefore, we repeat the procedure presented in L18 and we first correct for this effect by binning in metallicity the sample used by \citet{starkenburg17}, which provides both the SDSS spectroscopic metallicity and the Pristine photometric metallicity for several thousands stars. For each bin, the median difference between the SDSS and Pristine metallicities is computed. This procedure yields the bias as a function of the photometric metallicity, which is used to correct our whole Sgr~II metallicity catalog. All stars with [Fe/H]$_\mathrm{CaHK}$ $< -4.0$ or [Fe/H]$_\mathrm{CaHK}$ $> -1.0$ are discarded as our Pristine model is not reliable for those stars \citep{youakim17}. This choice does not affect the analysis as the systemic metallicity of Sgr~II is well within this range. Stars with a large uncertainty in the $CaHK$ photometry ($\delta_\mathrm{CaHK}  > 0.1$) are rejected. All remaining stars within 2$r_h$ are selected and their photometric metallicity distribution function (MDF) is shown in Figure \ref{histos_FeH} with the solid red line. The distribution of all field stars within $5r_h < r < 12r_h$ is shown as the black dashed line for comparison. 

Sgr~II stars in red stand out clearly in Figure \ref{histos_FeH} as they form a pronounced peak around [Fe/H]$_\mathrm{CaHK} \sim -2.3$ dex that does not exist in the MDF of the field stars in black.
To derive Sgr~II's metallicity properties, we assume that the population encapsulated inside 2r$_h$ (corresponding to 206 stars) in Figure \ref{histos_FeH} can be modelled as the sum of the foreground MDF and a normally distributed photometric metallicity population associated with Sgr~II stars. This assumption seems legitimate as the metallicity distribution at the metal-rich end of the red histogram in Figure \ref{histos_FeH} overlaps well with the black distribution, thus implying that the underlying foreground contamination MDF is comparable over the field of view. The Sgr~II stellar population metallicity distribution is assumed to be normally distributed, with a mean [Fe/H]$_\mathrm{CaHK}^\mathrm{SgrII}$ and a standard deviation of $\sigma = \sqrt{ (\delta[Fe/H]^\mathrm{CaHK}_k)^2 + (\sigma_\mathrm{[Fe/H]}^\mathrm{CaHK})^2 }$ with $\sigma_\mathrm{[Fe/H]}^\mathrm{CaHK}$ being the intrinsic metallicity dispersion of Sgr~II and $\delta[Fe/H]^\mathrm{CaHK}_k$ the uncertainty on the photometric metallicity of the $k$-th star. The likelihood model for the MW contamination stars is built by interpolating the [Fe/H]$_\mathrm{CaHK}$ MDF of all stars outside 5r$_h$. This model is then smoothed by a gaussian kernel of 0.2 dex to account for poor statistics in some metallicity bins. The fit is performed through a MCMC algorithm, and we marginalise over the foreground contamination model. At each iteration, we randomly draw a photometric metallicity for all stars in the contamination subsample, according to their individual photometric metallicity uncertainties. Then, the procedure to build the foreground contamination model described above is repeated. In doing so, the analysis takes into account the overall uncertainty of the contamination MDF. 

\begin{figure}
\begin{center}
\centerline{\includegraphics[width=\hsize]{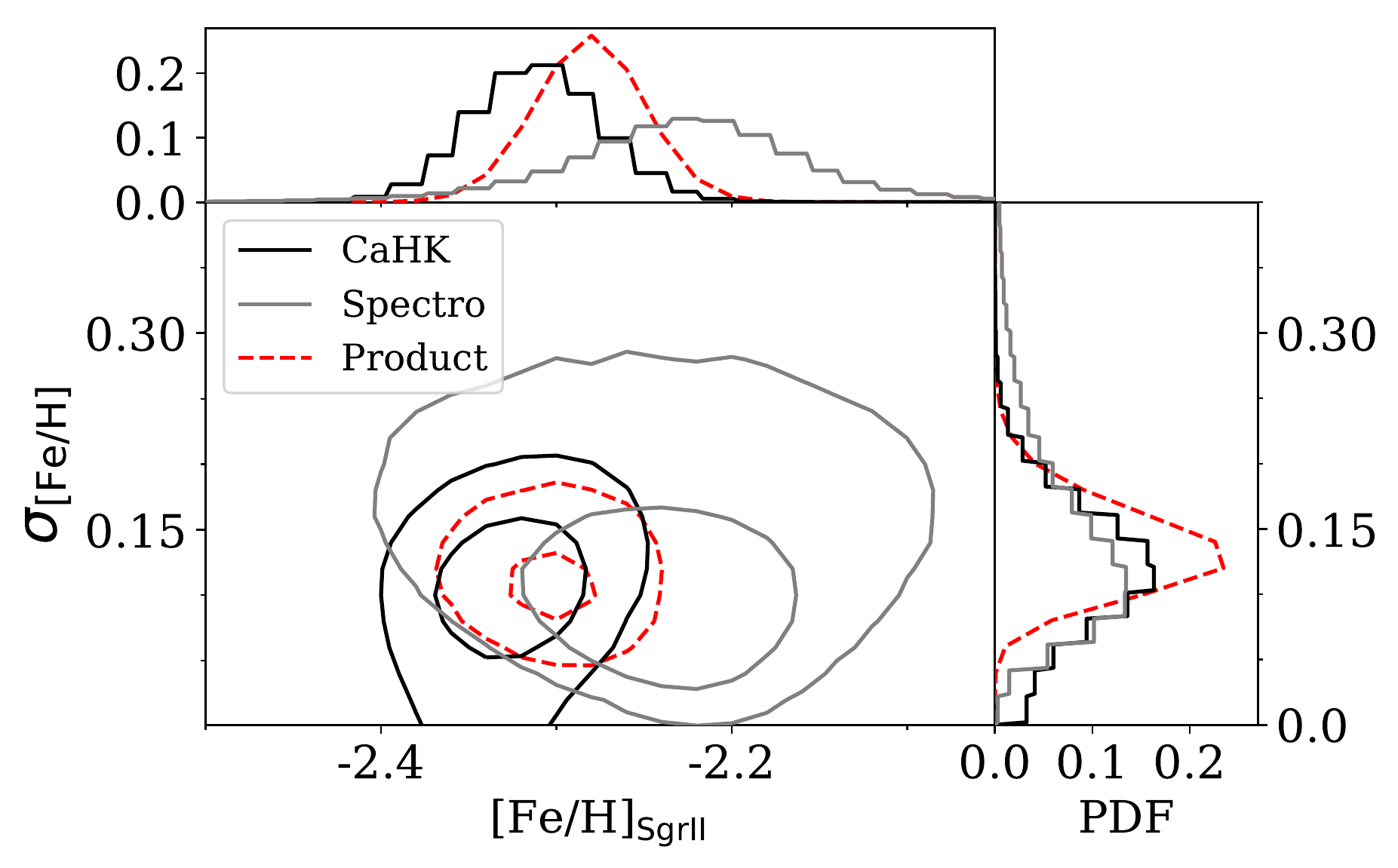}}
\caption{Two-dimensional joint PDFs of the systemic metallicity and dispersion for Sgr~II using the photometric CaHK metallicities (black) and the individual spectroscopic metallicities of member stars (grey). These two independent measurements are combined to give the final PDF shown as the dashed red line. The contours represent the 39, 88 and 95 \% volume intervals. The associated one-dimensional marginalised PDFs for all cases are shown in the upper and right panels. Both methods are in agreement and shows that Sgr~II has a small but measurable metallicity dispersion.}
\label{contours} 
\end{center}
\end{figure}

The 39, 88 and 95 \% volume intervals on the final 2D posterior PDF, corresponding to the $1$,$2$ and $3\sigma$ confidence levels for the 2D gaussian case, are shown in black solid line in Figure \ref{contours}.  We measure a significant, non-zero metallicity spread in Sgr~II, with $\sigma_\mathrm{[Fe/H]}^\mathrm{sgr} = 0.11^{+0.05}_{-0.03}$ dex and to be particularly metal-poor ([Fe/H]$_\mathrm{CaHK}^\mathrm{SgrII} = -2.32 \pm 0.04$ dex), in agreement with the stellar population inferred through the CMD fitting. To ensure that this inference is not caused by any systematic effect, the same analysis was done in L18 with two metal-poor globular clusters in the Pristine footprint, M92 ans M15. The systemic metallicities of both clusters were found to be compatible with their previous estimates using spectroscopic data. Furthermore, their metallicity dispersions were unresolved, in agreement with previous studies, showing that the technique does not seem to be affected by a systematic effect. \\

\section{Spectroscopic analysis}

\subsection{Velocity properties}

\begin{figure}
\begin{center}
\centerline{\includegraphics[width=\hsize]{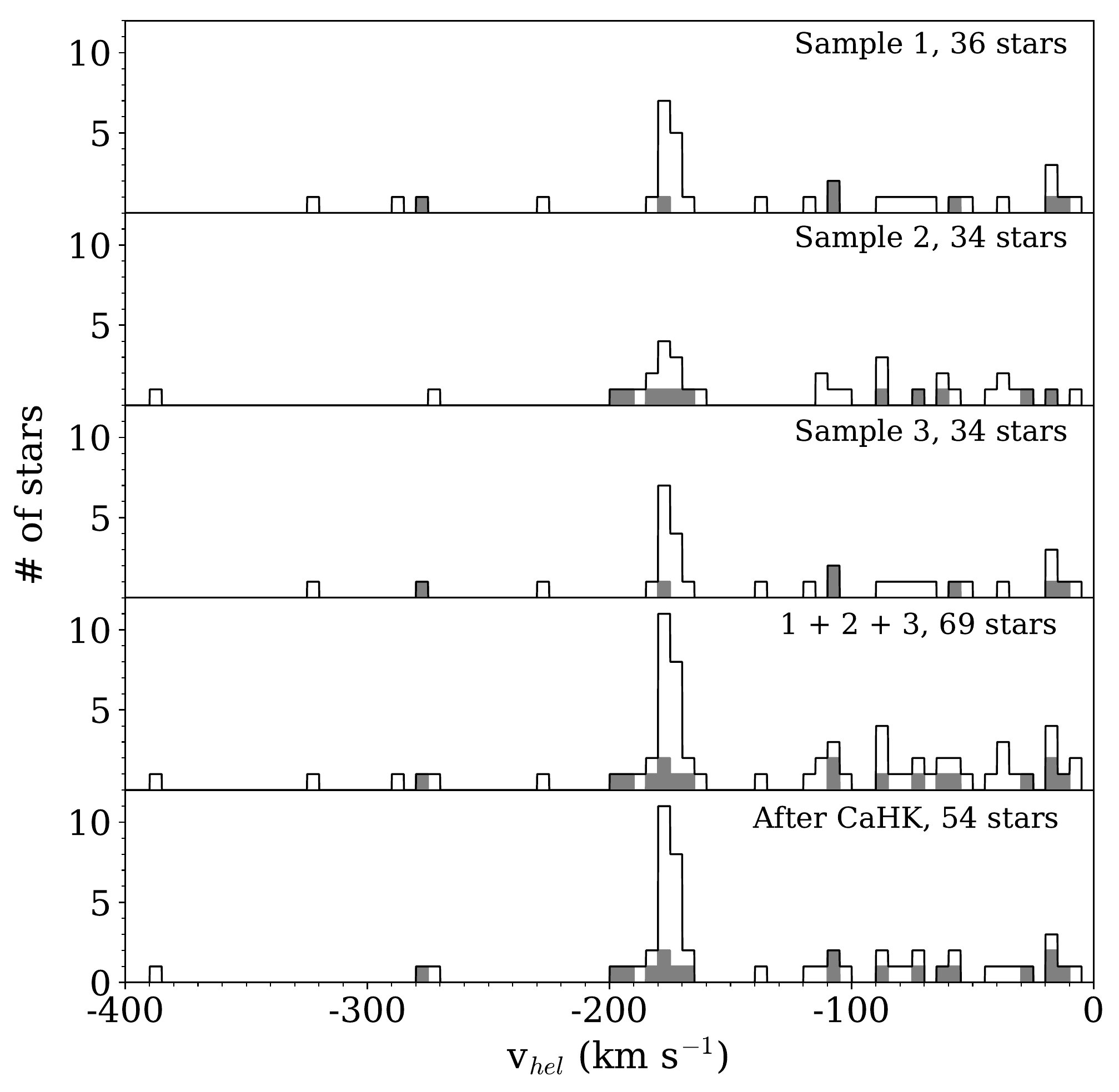}}
\caption{Heliocentric velocity histograms for the three spectroscopic samples. The fourth panel shows the merging of all samples. The grey histograms show the number of stars with a non-reliable photometric metallicity measurement in our sample that therefore cannot be filtered out by our technique. The peak of Sgr~II stars around $-177 \kms$ is pronounced and the disc contamination, from 0 to $-100$ $\kms$, is also quite populated. The last panel shows the final spectroscopic catalog filtered from ``metal-rich'' stars using photometric metallicities based on our CaHK photometry. For stars with reliable photometric metallicities, the ones with $-4.0 < $ [Fe/H]$_\mathrm{CaHK} < -1.6$, i.e. compatible with Sgr~II metallicity properties measured in section 3 and 5, are selected, while the others are discarded. Stars with mediocre quality CaHK measurement or [Fe/H]$_\mathrm{CaHK}$ uncertainties are not discarded as their [Fe/H]$_\mathrm{CaHK}$ is not reliable.  The disc and halo contamination is notably reduced, and one star located in the Sgr~II velocity peak is identified as a contaminant and removed from the sample.}
\label{histos_vel} 
\end{center}
\end{figure}

\begin{figure}
\begin{center}
\centerline{\includegraphics[width=\hsize]{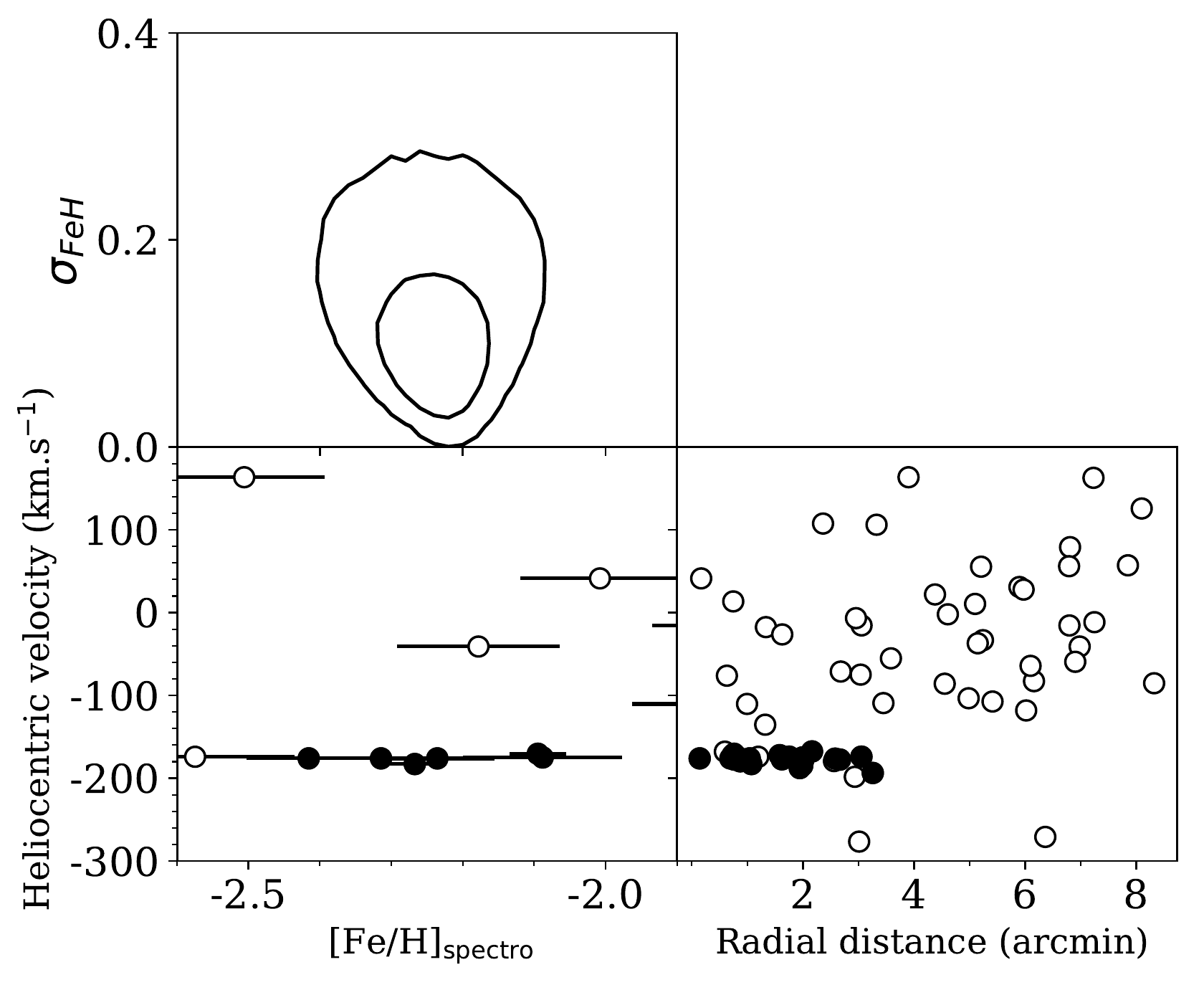}}
\caption{\textit{Bottom right panel:} Radial distances to the center of Sgr~II with respect to the heliocentric velocities. Open circles represent stars that are metal-poor using their photometric metallicities, or stars with non-reliable photometric metallicity measurements. Black-filled dots represent member stars. \textit{Bottom left panel:} Spectroscopic metallicities with respect to the heliocentric velocities of all stars from the final spectroscopic data set with S/N $>=$ 12. \textit{Top panel:} Two-dimensional joint PDF of the systemic spectroscopic metallicity and metallicity dispersion. The contours represent the local 39\%, 88\% and 95\% volume intervals. Sgr~II comes out as a very metal-poor satellite, with [Fe/H]$_\mathrm{spectro} = -2.23 \pm 0.05$ dex, and seems chemically enriched: $\sigma_\mathrm{[Fe/H]}^\mathrm{spectro} =  0.10 ^{+0.06}_{-0.04} $ dex.  }
\label{vel_vs_r} 
\end{center}
\end{figure}

\begin{figure}
\begin{center}
\centerline{\includegraphics[width=\hsize]{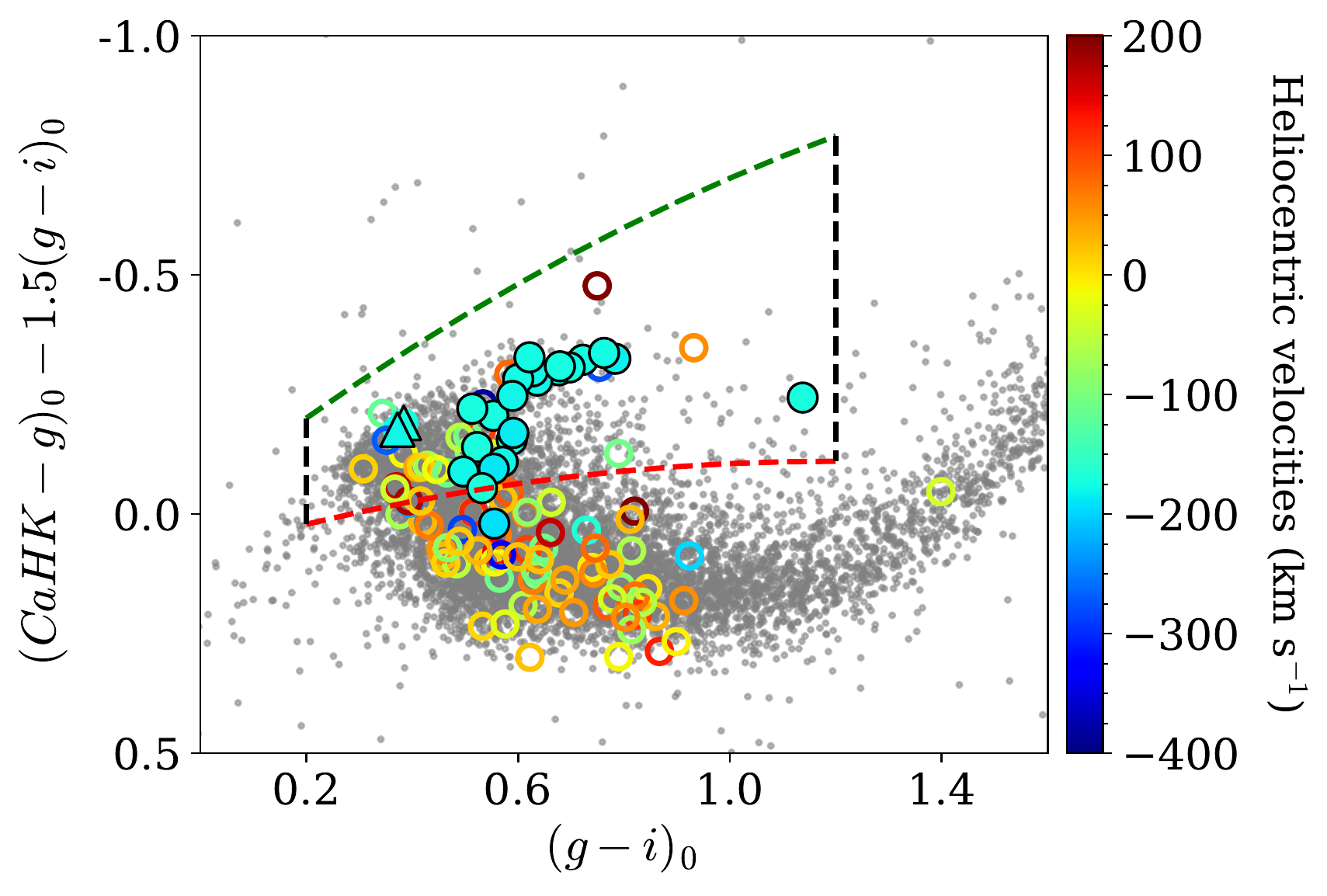}}
\caption{Pristine colour-colour diagram: the $(g-i)_0$ colour is represented on the x-axis, while the metallicity information is carried by the $(CaHK-g)_0 - 1.5*(g-i)_0$ colour on the y-axis. Grey dots stand for all field stars in a range between 5 and 8 half-light radii to Sgr~II. Most of them are halo and disc stars and form a stellar locus of more metal-rich stars ([Fe/H]$_\mathrm{CaHK}$ $\sim$ 1 or above). Stars observed with spectroscopy are represented with circles colour-coded according to their heliocentric velocities. Among those, filled circles show the stars identified as spectroscopic members, while filled triangles stand for the HB stars in the spectroscopic data set. Above the grey stellar locus are located stars that become more and more metal-poor as we go towards the upper part of the diagram. Two iso-metallicity sequences are shown in red and green dashed lines, corresponding respectively to a photometric metallicity of [Fe/H]$_\mathrm{CaHK}$ $\sim$ -1.6 and [Fe/H]$_\mathrm{CaHK}$ $\sim$ -4.0. As expected, most of the stars in cyan, with a radial velocity compatible with Sgr~II, are located in the metal-poor region enclaved by those two sequences. Hence, only stars within this region are selected for the final spectroscopic sample. Furthermore, we add a criteria on $(g-i)_0$ and discard all stars between $0.2 < (g-i)_0 < 1.2$ in order to discard potential foreground white dwarfs.}
\label{cahk} 
\end{center}
\end{figure}

\begin{figure}
\begin{center}
\centerline{\includegraphics[width=\hsize]{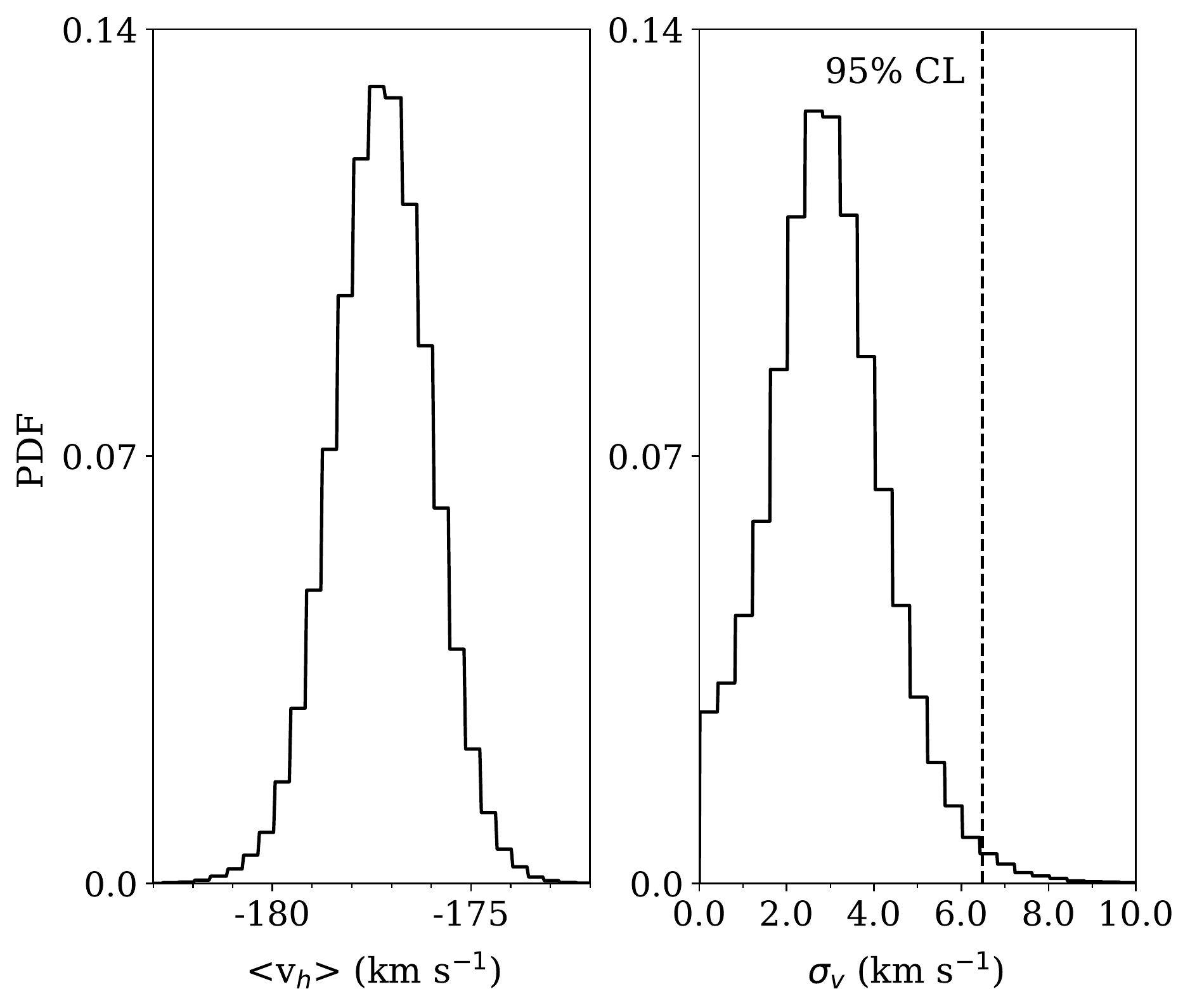}}
\caption{ Marginalised PDFs for the systemic velocity (left) and its associated dispersion (right) for Sgr~II. The satellite is found to be dynamically cold: the velocity dispersion is only marginally resolved: $\sigma_{vr} = 2.7^{+1.3}_{-1.0} \kms$, reaching 6.5 $\kms$ at the 95\% confidence level.}
\label{vel} 
\end{center}
\end{figure}

The systemic velocity and velocity dispersion are derived using the deep spectroscopic observations of the system. The heliocentric velocity distributions of each mask separately are summarised in the top three panels of Figure \ref{histos_vel}, and are combined to obtain one single velocity distribution shown in the fourth panel of the same figure. We present the radial distance of each star with respect to their radial velocities in Figure \ref{vel_vs_r}. The full dataset is detailed in Table 2.

The velocity peak of Sgr~II stands out at around $-177$ km s$^{-1}$ while contaminating MW stars are distributed sparsely all over the velocity space, and in particular can be located in the vicinity of the Sgr~II velocity peak. Because of the small number of stars in the Sgr~II population, the velocity properties can be polluted by the contamination. Ideally, those stars would have to be identified and discarded from the spectroscopic sample. Sgr~II is an old and metal-poor system as suggested by its CMD, and confirmed in section 3, 4 and 5, whereas the contaminating foreground is expected to be more metal-rich overall. Therefore, the contamination stars could be discarded based on their metallicities. Even though the individual [Fe/H] are only accessible for the brightest stars in our sample in our spectroscopic sample, the Pristine $CaHK$ photometry can be used here. 

The Pristine colour-colour diagram is shown in Figure \ref{cahk}. Field stars, i.e. a randomly selected sample of all stars outside five half-light radii, are represented in small black dots and form a clear stellar locus. This diagram is constructed so that the individual metallicity of a given star decreases from the bottom right to the top left, while more metal-rich stars occupy the lower part of this colour-colour space. Stars observed with spectroscopy are colour-coded according to their heliocentric velocities, provided they pass the following CaHK photometry and [Fe/H]$_\mathrm{CaHK}$ quality cuts: $\Delta [Fe/H]_\mathrm{CaHK} < 0.5 $ and $\delta_\mathrm{CaHK} < 0.1 $. Stars that do not match these criteria are not discarded from the final spectroscopic sample, whether they are more metal-rich or not in the model, as our goal here is to select only stars for which the CaHK photometry is reliable enough to ensure their metal-poor nature. In section 4, we found that Sagittarius II has a systemic metallicity of [Fe/H]$_\mathrm{CaHK}^\mathrm{SgrII} = -2.32 \pm 0.04$ dex and has a resolved metallicity dispersion. Therefore, within the subsample of stars that passed the CaHK photometry cuts discussed above, we choose to select only stars with $ -4.0 < [Fe/H]_\mathrm{CaHK} < -1.6 $, as a Sgr~II-like system would likely have a star formation history too short to produce significantly more metal-rich stars. The region of the diagram that corresponds to such a metallicity cut is represented by the two iso-metallicity sequences in green and red dashed lines in Figure \ref{cahk}. Two cuts in $(g - i)_0$ are also applied in order to discard potential white dwarfs and metal-rich stars. The final spectroscopic velocity distribution is shown in the last panel of Figure \ref{histos_vel}. A significant number of MW stars with a reliable Pristine photometric metallicity measurements are cleaned out from the catalog as their metallicities are too high to be members of Sgr~II, even if the satellite has a metallicity spread. In particular, one star in the immediate vicinity of the Sgr~II velocity peak is identified as a more metal-rich contaminants using this technique ($\FeH_{\mathrm{CaHK}} = -1.11 \pm 0.25$ dex) and therefore discarded. 

The resulting velocity distribution is assumed to be the sum of three normally distributed populations:  one for Sgr~II stars, and two others corresponding to the MW foreground disc and halo stars. Each individual likelihood is weighted by its CMD and structural probability membership determined in section 5.  One can write the individual likelihood of the $k$-th star as

\begin{eqnarray}
\mathcal{L}(\langle \mathrm{v}_\mathrm{SgrII} \rangle,\sigma_v,\langle \mathrm{v}_\mathrm{MWd} \rangle,\sigma_\mathrm{vd},\langle \mathrm{v}_\mathrm{MWh} \rangle,\sigma_\mathrm{vh}   | \{\mathrm{v}_{\mathrm{r},k},\delta_{\mathrm{v},k}\}) =  \\ \nonumber
\prod_k ( (1 - \eta_\mathrm{MWd} - \eta_\mathrm{MWh}) P_\mathrm{mem} G(\{ \mathrm{v}_{\mathrm{r},k} \} | \langle \mathrm{v}_\mathrm{SgrII} \rangle,\sigma_v)    \\ \nonumber 
+ \; (1 - P_\mathrm{mem}) ( \eta_\mathrm{MWd} G(\{ \mathrm{v}_{\mathrm{r},k} \} | \langle \mathrm{v}_\mathrm{MWd} \rangle,\sigma_\mathrm{vd})   \\ \nonumber
+ \; \eta_\mathrm{MWh} G(\{ \mathrm{v}_{\mathrm{r},k} \} | \langle \mathrm{v}_\mathrm{MWh} \rangle,\sigma_\mathrm{vh}))), \nonumber
\end{eqnarray}

\noindent with $\sigma_v = \sqrt{ (\sigma^{\mathrm{SgrII}}_\mathrm{v})^2 + \delta_{\mathrm{v},k}^2 + \delta_{thr,i} }$ and $\delta_{\mathrm{v},k}$ the individual velocity uncertainty of the k-th star, $ \sigma^{\mathrm{SgrII}}_\mathrm{v}$ the intrinsic velocity dispersion, $\delta_{thr}$ the systematic threshold derived in section 2.2. $\langle \mathrm{v}_\mathrm{SgrII} \rangle$ is the systemic velocity of Sgr~II. $\eta_\mathrm{MWd}$ and $\eta_\mathrm{MWh}$ are the fractions of stars respectively in the MW disc and halo populations. $\sigma_\mathrm{vd}$ is defined as $\sigma_\mathrm{vd} = \sqrt{ (\sigma^{\mathrm{MWd}}_\mathrm{v})^2 + \delta_{\mathrm{v},k}^2 + \delta_{thr,i} }$, with $\sigma^{\mathrm{MWd}}_\mathrm{v}$ the intrinsic velocity dispersion of the disc population. The corresponding quantity for the halo population is written $\sigma_\mathrm{vh}$, while $\langle v_\mathrm{MWd} \rangle$ is the systemic velocity of the disc population in the sample (resp. for the halo population). $G$ is the usual one-dimensional normal distribution. We run a MCMC analysis and show the resulting marginalised 1D PDFs in Figure \ref{vel}.  At each iteration of the MCMC,  the systematic threshold $\delta_\mathrm{thr}$ is randomly drawn from its PDF. The favoured systemic velocity is $\langle \mathrm{v}_\mathrm{SgrII} \rangle = -177.3 \pm 1.2$ km s$^{-1}$. The velocity dispersion of Sgr~II is $\sigma^{\mathrm{SgrII}}_v = 2.7^{+1.3}_{-1.0} $ km s$^{-1}$, reaching 6.5 km s$^{-1}$ at the 95\% confidence interval, thus showing that Sagittarius II is a dynamically cold satellite. A similar analysis was performed for the inner (r $<$ 1 arcmin) and outer (r $>$ 1 arcmin) regions and no statistical difference in terms of velocity dispersion was found.

\subsection{Metallicity properties}

To infer the metallicity properties of Sgr~II from the spectrocopy, we create a subsample constituted of stars brighter than $i_0 = 20.5$ and a S/N ratio above 12 from our final spectroscopic sample, for a total of 26 stars. The spectroscopic metallicity is estimated using the calibration from \citet{starkenburg10} based on the Ca triplet. This method is originally calibrated for RGB stars, however, \citet{leaman13} showed that it can applied to stars up to two magnitudes fainter (see also \citealt{carrera13}). The resulting sample consists of six member stars, for which the individual spectroscopic metallicities are reported in Table 2 under ``Fe/H]$_\mathrm{spectro}$''. The distribution of spectroscopic metallicities with respect to the radial velocity is shown in the bottom left panel of Figure \ref{vel_vs_r}, and shows the existence of a clump of stars at around $\FeH_\mathrm{spectro} \sim -2.3$ dex at the velocity of Sgr~II.

To derive the systemic metallicity and metallicity dispersion of Sgr~II, we assume that the spectroscopic metallicity of Sgr~II stars are normally distributed and weigh each star with its CMD and structural probability membership, giving the following likelihood function

\begin{eqnarray}
\mathcal{L}(\langle \FeH_\mathrm{spectro} \rangle,\sigma_\mathrm{[Fe/H]} | \{\FeH_{\mathrm{spectro},k},\delta_{\mathrm{[Fe/H]},k}) = \\
P_\mathrm{mem}  \; G(\FeH_{\mathrm{spectro},k},\delta_{\mathrm{[Fe/H]},k} | \langle \FeH_\mathrm{spectro} \rangle,\sigma_\mathrm{[Fe/H]} \}) \nonumber
\end{eqnarray}

\noindent with $\sigma_\mathrm{[Fe/H]} = \sqrt{\delta_{\mathrm{[Fe/H]},k}^2 + (\sigma_\mathrm{[Fe/H]}^{sgr})^{2}}$, $\delta_{\mathrm{[Fe/H]},k}$ being the individual uncertainty on the spectroscopic metallicity of the $k$-th star, and $\sigma_\mathrm{[Fe/H]}^{sgr}$ the intrisic metallicity dispersion of Sagittarius II.  The 39, 88 and and 95 $\%$ volume intervals are represented by black solid lines on the resulting 2D probability distribution functions (PDF) in the top left panel of Figure \ref{vel_vs_r}. Sgr~II is confirmed to be metal-poor, with [Fe/H]$_\mathrm{spectro}^\mathrm{SgrII}$ = -2.23 $\pm$ 0.05 dex. Moreover, we find a metallicity dispersion of $\sigma_\mathrm{spectro}^{[Fe/H]} = 0.10 ^{+0.06}_{-0.04} $ dex. This spread in metallicity is driven by the two brightest stars identified as members of Sgr~II, for which the spectroscopic metallicity is accurately measured. They have respectively a spectroscopic metallicity of $-2.27 \pm 0.04$ dex and $-2.10 \pm 0.04$ dex. Furthermore, since they are among the stars that were observed multiple times in our catalog, it is possible to infer their individual spectroscopic metallicities using the Ca triplet equivalent widths of each run separately. For both stars, the metallicities obtained from each spectroscopic run in which they were observed are consistent with one another, suggesting that their final $\FeH$ are not driven by one spurious equivalent widths measurement in one of the three spectroscopic samples. In addition with being consistent with the CMD of Sgr~II and its systemic velocity, the two stars are also remarkably compatible with the satellite's proper motion inferred in section 6. Taken all together, we favour the fact that these two stars are indeed members of Sgr~II, and there is more than one stellar population in the system.

The two independent measurements of the metallicity and dispersion of the satellite, using the $CaHK$ observations on the one hand and the spectra on the other, are perfectly compatible. The results of both methods are then combined into one single measurement by performing the product of the two 2D joint PDFs. We show the corresponding 39, 88 and 95 \% volume intervals in red thick line in Figure \ref{contours}. This final measurement yields a systemic metallicity of [Fe/H]$_\mathrm{SgrII}$ = -2.28 $\pm$ 0.03 dex and a metallicity dispersion of $\sigma_\mathrm{[Fe/H]}^{SgrII} = 0.12 ^{+0.03}_{-0.02}$ dex.

\section{\textit{Gaia} DR2 proper motions and orbit}

\begin{figure}
\begin{center}
\centerline{\includegraphics[scale=0.45]{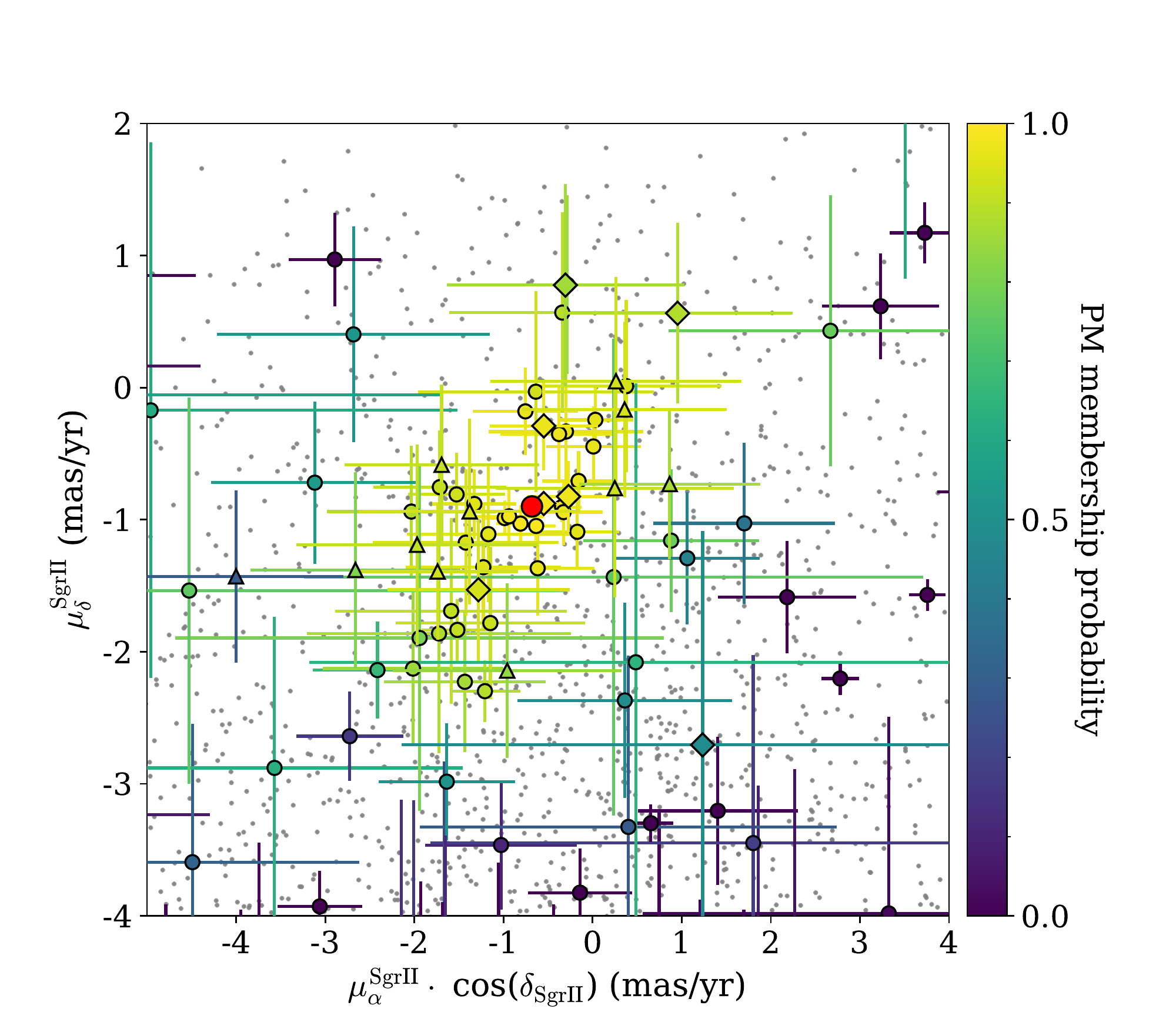}}
\caption{Distribution of the proper motions of MW contamination stars in small grey dots, and our Sgr~II-like population shown with dots colour-coded according to their proper motion membership probability, derived from a gaussian mixture model. The seven Sgr~II members with a proper motion measurement in \textit{Gaia} are represented with diamonds and the twelve HB stars with triangles. The systemic proper motion of Sgr~II ($\mu_{\alpha}^{*} = -0.65^{+0.08}_{-0.10}$ mas.yr$^{-1}$, $\mu_{\delta} = -0.88 \pm 0.12$ mas.yr$^{-1}$) is represented with a large red dot.}
\label{proper_motions}
\end{center}
\end{figure}

\begin{figure*}
\begin{center}
\centerline{\includegraphics[scale=0.5]{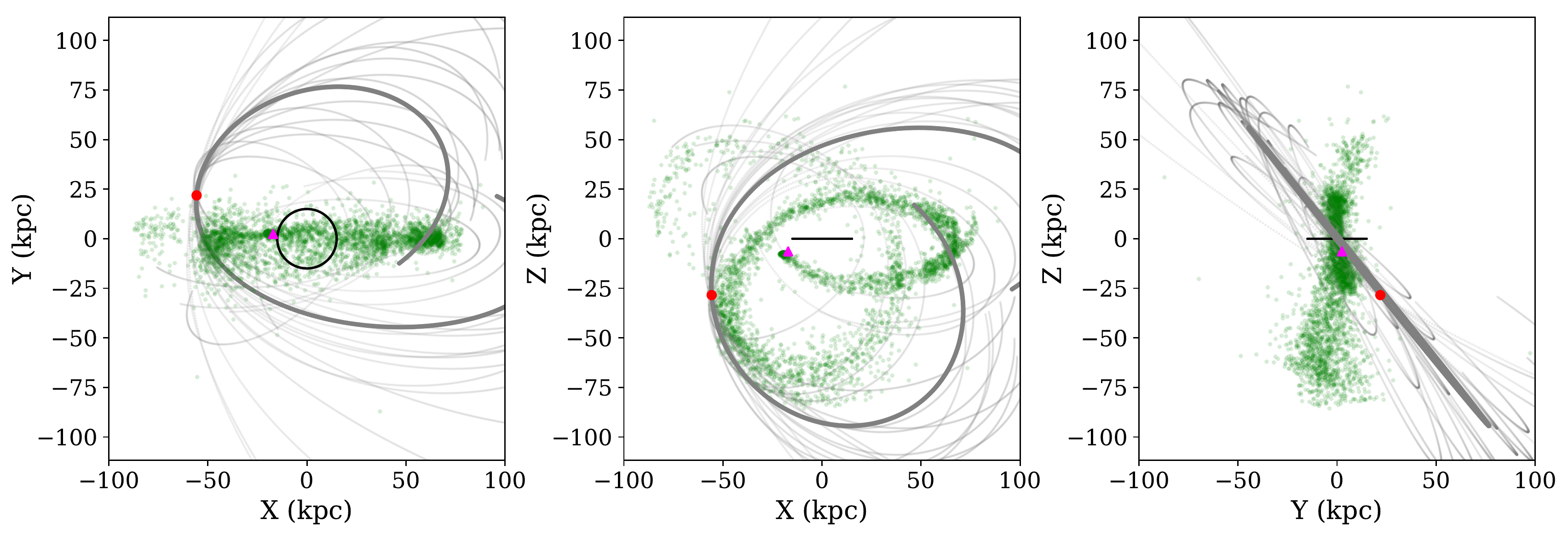}}
\caption{Projections of the orbit of Sgr~II on the X-Y, X-Z and Y-Z planes backwards and forwards over 2.0 Gyr. Twenty-one orbits are shown here: the one based on the favoured position, distance, radial velocity and PMs of the satellite (as the thick, darker grey line), and twenty others using random realisations of those parameters (as thin, slightly transparent grey lines). The red circle is the current position of Sgr~II, while the magenta triangle is the one of the Sgr dSph. A N-body simulation of the trailing arm of the Sgr stream \citep{law_majewski10} is shown in green. The MW disk is shown in black, with a chosen radius of 15 kpc.}
\label{orbit}
\end{center}
\end{figure*}

To infer the orbit of Sgr~II, we first build a sample of Sgr~II-like population based on the mask shown in the right panel of Figure \ref{CMDs}. The proper motions of those stars are retrieved from the \textit{Gaia} Data Release 2 \citep{brown18}.  All member stars identified with spectroscopy and bright enough to have a proper motion measurement in \textit{Gaia} are naturally present in this sample. Furthermore, the \textit{Gaia} DR2 data are also cross-matched with the potential HB stars within two half-light radii of the satellite. Twelve HB stars have a proper motion measurement in \textit{Gaia} and are added to the sample shown in Figure \ref{proper_motions}.

The inference of the Sgr~II proper motion is performed with a gaussian mixture model. We assume that the sample can be modelled by the sum of two bivariate gaussians: one for the Sgr~II population and another for the foreground MW contamination. The sets of parameters inferred from the analysis are composed of the proper motions in both directions, their dispersions and correlation $c$, for Sgr~II ($\mathcal{P}_\mathrm{SgrII} = \langle \mu_{\alpha,\mathrm{SgrII}}^{*} \rangle$,$ \langle \mu_{\delta,\mathrm{SgrII}} \rangle$,$\sigma_\mathrm{1}$,$\sigma_\mathrm{2}$,$c_\mathrm{SgrII}$) and for the contamination ($\mathcal{P}_\mathrm{MW} = \langle \mu_{\alpha,\mathrm{MW}}^{*} \rangle$,$ \langle \mu_{\delta,\mathrm{MW}} \rangle$,$\sigma_\mathrm{3}$,$\sigma_\mathrm{4}$,$c_\mathrm{MW}$). The proper motion properties of the $k$-th star are defined as $\vec{d_{k}} = \{\mu_{\alpha,k}^{*},\mu_{\delta,k},\delta\mu_{\alpha,k}^{*},\delta\mu_{\delta,k}\}$ with $\delta\mu_{\alpha,k}^{*}$ the uncertainty on the proper motion in the RA direction (respectively for DEC). The individual likelihood is

\begin{eqnarray}
\mathcal{L}(\mathcal{P}_\mathrm{SgrII},\mathcal{P}_\mathrm{MW} | \vec{d_{k}})  =  \prod_k \eta P_{mem} \mathcal{M_G}(\vec{d_{k}} | \mathcal{P}_\mathrm{SgrII},\mathcal{P}_\mathrm{MW})    \\ \nonumber 
+ \;  ( 1 - \eta )  (1 - P_{mem}) \mathcal{M_G}(\vec{d_{k}} | \mathcal{P}_\mathrm{SgrII},\mathcal{P}_\mathrm{MW}),   \\ \nonumber
\end{eqnarray}

\noindent where $\mathcal{M_G}$ is a two-dimensional gaussian and $\eta$ the fraction of Sgr~II stars in the sample.

The gaussian mixture model gives a systemic proper motion of $\mu_{\alpha}^\mathrm{*,SgrII} = -0.65^{+0.08}_{-0.10}$ mas yr$^{-1}$ and $\mu_{\delta}^\mathrm{SgrII} = -0.88 \pm 0.12$ mas yr$^{-1}$ for Sgr~II. These proper motions take into account the systematic error on the one for dSph derived by \textit{Gaia} Collaboration et al. (2018b). We also inferred the proper motion of the system using the HB and spectroscopic member stars only, and found a compatible result with ($\mu_{\alpha}^{*}$,$\mu_{\delta}$) = ($-0.55 \pm 0.13$,$-0.80 \pm 0.08$) mas yr$^{-1}$. 

Our estimate is discrepant from the one of \citet{massari18} who find a proper motion of ($\mu_{\alpha}^{*}$,$\mu_{\delta}$) = ($-1.18\pm 0.14$,$-1.14 \pm 0.11$) mas yr$^{-1}$. They rely on the convergence of the astrometric parameters through a $2.5\sigma$ clipping procedure, with an initial guess on those parameters based on the potential HB stars of Sgr~II. However, our measurement based only on HB and spectroscopic member stars gives credit to the proper motion found in this work, and disfavours the estimate of the work of \citet{massari18}, which might be biased by the foreground contamination.

The orbit of the satellite can then be inferred using the GALPY package \citep{bovy15}. The MW potential chosen to integrate the orbit is a modified ``MWPotential14'' constituted of three main components: a power-law, exponentially cut-off bulge, a Miyamoto-Nagai Disc, and a NFW DM halo with a virial mass of $1.2 \times 10^{12} \msun$. Further details about this MW potential model can be found in \citet{bovy15}. We integrate 2000 orbits backwards and forwards, each time by randomly drawing a position, distance, radial velocity and proper motions from their respective PDFs, and extract for each realisation the pericenter, apocenter and ellipticity. Each orbit is shot over 2 Gyr. The favoured orbit (i.e. the favoured position, distance, radial velocity and PMs) is shown in Figure \ref{orbit} in the X-Y, X-Z and Y-Z planes, along with the Sgr stream. Twenty other random realisations of Sgr~II orbits are also shown in grey, partially transparent lines.

The analysis yields a pericenter of $54.8 ^{+3.3}_{-6.1} \kpc$, an apocenter of $118.4 ^{+28.4}_{-23.7} \kpc$ and an orbital ellipticity of $0.44 \pm 0.01$. Moreover, Figure \ref{orbit} shows that the orbit of Sgr~II is compatible with the trailing arm of the Sgr stream, despite being slightly tilted from it, especially in the Y direction.

\section{Discussion}

\begin{figure*}
\begin{center}
\centerline{\includegraphics[width=\hsize,scale=0.8]{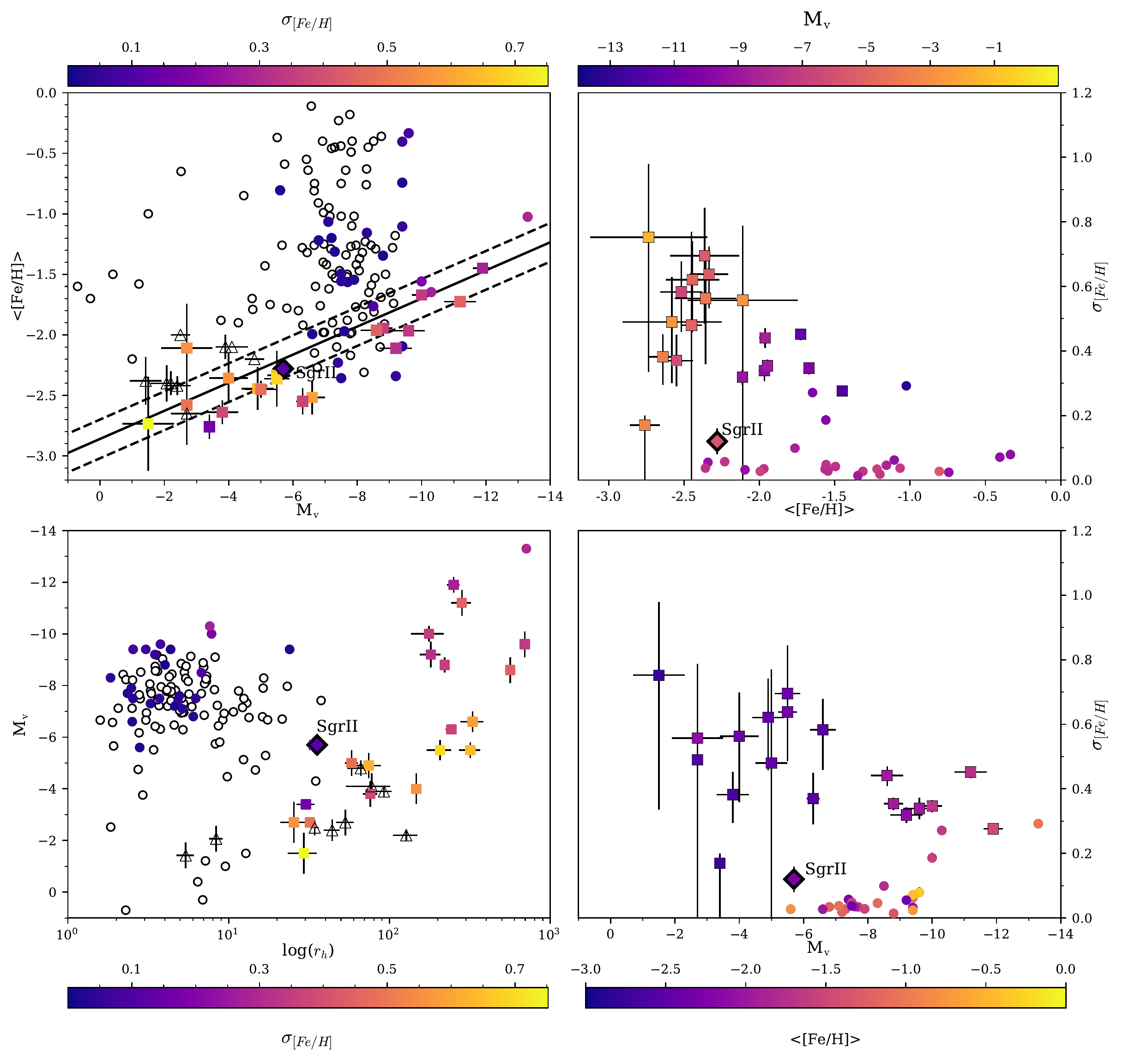}}
\caption{ Comparison of Sgr~II with other GCs and dwarf galaxies of the Milky Way. Squares represent dwarf galaxies while circles represent globular clusters, and the diamond corresponds to Dra~II. Triangles stand for recently discovered dwarf-galaxy candidates that await confirmation. Hollow markers correspond to systems for which no metallicity dispersion measurement can be found in the literature. The solid line in the top-left panel corresponds to the luminosity-metallicity relation of \citet{kirby13} for dwarf spheroidals and dwarf irregulars. Dashed lines represent the RMS about this relation, also taken from \citet{kirby13}. Among the 123 globular clusters presented here, the properties of 116  were extracted from \citet{harris96} catalog, revised in 2010. For the remaining ones (Kim 1, Kim 2, Kim 3, Laevens 1, Balbinot 1, Munoz 1 and SMASH 1) parameters of the discovery publications were used (\citet{kim15b}, \citet{kim15}, \citet{kim16b}, \citet{laevens14}, \citet{balbinot13}, \citet{munoz12b} and \citet{martin16b}). Globular cluster metallicity spread measurements are taken from \citet{willman_strader12} and references therein: \citet{carretta06,carretta07b,carretta09b,carretta11}, \citet{cohen10}, \citet{gratton07}, \citet{johnson_pilachowski10}, and \citet{marino11}. \citet{mcconnachie12} and \citet{willman_strader12} are used to compile the properties of the dwarf galaxies represented here. The 18 dwarf galaxies represented here are: Bootes I \citep{belokurov06,norris10}, Canes Venatici I \citep{zucker06b}, Canes Venatici II \citep{sakamoto06}, Coma Berinices, Hercules, Leo IV and Segue I \citep{belokurov07}, Draco and Ursa Minor \citep{wilson55}, Fornax \citep{shapley38b}, Leo I and Leo II \citep{harrington_wilson50}, Pisces II \citep{belokurov10}, Sculptor \citep{shapley38a}, Sextans \citep{irwin90}, Ursa Major I \citep{willman05b}, Ursa Major II \citep{zucker06a}, Willman I \citep{willman05a}. Their metallicity and metallicity spreads were drawn from \citet{kirby08}, \citet{kirby10}, \citet{norris10}, \citet{willman11}. The dwarf galaxy candidates discovered recently and shown on this figure are Bootes II \citep{koch_rich14}, DES1 \citep{luque16,conn18}, Eridanus III \citep{bechtol15,conn18,koposov15}, Hyades II \citep{martin15}, Pegasus III \citep{kim15b}, Reticulum II and Horologium I \citep{koposov15b}, Segue II \citep{belokurov09}, and the most significant candidates of \citet{drlica-wagner15}: Gru II, Tuc III, and Tuc IV.}
\label{context} 
\end{center}
\end{figure*}

We used deep MegaCam broadband photometry, the narrow-band $CaHK$ Pristine survey and DEIMOS spectroscopy to conduct a thorough study of the Milky Way satellite Sgr~II. By performing a CMD and structural analysis, the satellite is found to have a size of 35.5 $^{+1.4}_{-1.2}$ pc, and is located at $73.1^{+1.1}_{-0.7}$ kpc based on the combination of BHB stars distances and a CMD fitting procedure. The favoured stellar population is old ($12.0 \pm 0.5$ Gyr) and metal-poor. Using our spectroscopic catalog, we are able to find the systemic velocity of Sgr~II to be $\langle \mathrm{v}_\mathrm{SgrII} \rangle =  - 177.3 \pm 1.3 \kms$. The velocity dispersion yields $\sigma^{\mathrm{sgr}}_\mathrm{v} = 2.7^{+1.3}_{-1.0} $ km s$^{-1}$ and is $ < 6.5 \kms$ at the 95\% confidence interval. From this spectroscopic analysis, 22 stars are identified as members of the satellite and reported in Table 2. The individual photometric metallicities provided by the Pristine survey are used to show that Sgr~II is a very metal-poor system, with [Fe/H]$_\mathrm{CaHK}^\mathrm{SgrII} = -2.32 \pm 0.04$ dex, and has a resolved metallicity dispersion: $\sigma_\mathrm{[Fe/H]}^\mathrm{CaHK} = 0.11^{+0.05}_{-0.03}$ dex. These two chemical properties are perfectly supported by our spectroscopic analysis of six RGB stars. We applied the Ca triplet calibration from \citet{starkenburg10} to derive Sgr~II spectroscopic metallicity and its associated dispersion: [Fe/H]$_\mathrm{spectro}^\mathrm{SgrII} = -2.23 \pm -0.08$ dex and $\sigma_\mathrm{[Fe/H]}^\mathrm{CaHK} = 0.10 ^{+0.06}_{-0.04}$ dex. Combining the CaHK and spectroscopic measurements, we obtain refined estimates of both parameters: [Fe/H]$_\mathrm{SgrII}$ = $-2.28 \pm 0.04$ dex and $\sigma_\mathrm{[Fe/H]}^\mathrm{SgrII} = 0.12^{+0.03}_{-0.02}$ dex. Finally, using the \textit{Gaia} DR2 data, the proper motion of Sgr~II is inferred to be ($\mu_{\alpha}^{*}$,$\mu_{\delta}$) = ($-0.65^{+0.08}_{-0.10}$,$-0.88 \pm 0.12$) mas yr$^{-1}$. This yields an apocenter and pericenter of $118.4^{+28.4}_{-23.7}$ and $54.8^{+3.3}_{-6.1}$ kpc respectively.

Sgr~II is in perfect agreement with the luminosity-metallicity relations for dwarf galaxies \citep{kirby13}, as it is shown in the top left panel of Figure \ref{context}. Sgr~II is however somewhat of an outlier in the $r_h$--$M_V$ plane (bottom-left panel of Figure \ref{context}), which led M18 to conclude that Sgr~II is a globular cluster. However, the locus of dwarf galaxies in this plane becomes uncertain at low luminosities. The satellite is still more extended than the vast majority of MW globular clusters as shown in the bottom left panel of Figure \ref{context}, although two of them have a comparable size: Crater \citep{belokurov14,laevens14} and Terzan~5 \citep{terzan68}. These two extended clusters do not, however, share the same metallicity properties as Sgr~II: Terzan~5 is a bulge cluster with [Fe/H] $> -0.5$ and Crater is more metal-rich with a systemic metallicity of [Fe/H] $ \sim -1.65$ \citep{weisz16}. Our two estimates of the metallicity dispersion of Sgr~II both yield similar results and suggest that the satellite was able to retain its gas and form successive generation of stars, thus suggesting the presence of a dark matter halo \citep{willman_strader12}. However, this result is driven by two bright RGB stars that have discrepant metallicity measurements. If one of the two were misidentified as a Sgr~II member, the claim of a metallicity dispersion would be weaker. 

The question of the dynamical mass of Sgr~II remains open. We can use the relation of \citet{walker09} to estimate the expected velocity dispersion of a purely baryonic system. Assuming a mass-to-light ratio of 2 for an old and metal-poor stellar population \citep{mclaughlin05}, Sgr~II would have a velocity dispersion of $\sim 1 \kms$, which is not incompatible with our inference of  $\sigma_{vr} =  2.7^{+1.3}_{-1.0} \kms$. Nevertheless, taken at face value, our velocity dispersion measurement implies that Sgr~II has a dynamical mass-to-light ratio of $9.5^{+14.8}_{-9.5} \msun \lsun^{-1}$ and favours a slightly DM-dominated system under the usual assumption of dynamical equilibrium and sphericity. If this is confirmed, it would mean that Sgr~II inhabits one of the lowest mass DM subhalo. Alternatively, this result could be driven by the compactness of the satellite, whose stars only probe the inner parts of the subhalo. 

Taken together, these two pieces of evidence (marginally resolved metallicity dispersion and plausibly non-baryonic M/L) would indicate that Sgr~II is more likely a dwarf galaxy rather than a cluster. 

Before the submission of this work, a spectroscopic study of Sgr~II was presented at the AAS iPoster\footnote{\url{<https://aas233-aas.ipostersessions.com/default.aspx?s=E7-10-7C-92-5D-B1-84-24-1F-B5-07-1A-BF-2E-10-65>}} session \citep{simon_fu19}. Using Magellan/IMACS spectroscopy, they found a systemic velocity and metallicity compatible to the ones in this work: $\langle v_r \rangle = -177.3 \pm 0.7 \kms$ and $\langle \FeH \rangle = -2.28$ dex. Their velocity dispersion is also consistent with ours: $\sigma_{vr} = 1.6 \pm 0.3 \kms$. Finally, the proper motion they derive for Sgr~II (($\mu_{\alpha}^{*}$,$\mu_{\delta}$) = ($-0.63^{+0.08}_{-0.10}$,$-0.89 \pm 0.06$) mas yr$^{-1}$) is also compatible with our work. However, they estimate a very low metallicity dispersion, with $\sigma_{\FeH} < 0.08$ dex at the 95 per cent confidence limit. Therefore, they conclude that the satellite is a globular cluster. Once the two data sets are made public, a thorough investigation is needed to understand the source of this discrepancy. Anyhow, it illustrates the difficulty of studying and understanding such faint systems.

Independently of the nature of Sgr~II, the orbit we infer for the satellite is compatible with the trailing arm of the Sagittarius stream according to the model by \citet{law_majewski10} (Figure \ref{orbit}). However, we note that the agreement between the two orbits is not perfect and, in particular, that the position of Sgr~II today and its favoured movement in the Y-Z galactocentric plane are slightly offset from the plane of the Sgr stream. Three hypotheses can be formulated to explain this discrepancy:

\begin{itemize}
\item The fact that the Sgr stream and the Sgr~II orbits are compatible is purely coincidental.
\item Sgr~II is linked to the stream, and the discrepancy between Sgr~II and the stream in the Y direction, if real, could be explained by the fact that Sgr dSph satellites were stripped first and with a different energy than that of stars represented in the simulation.
\item Sgr~II is linked to the stream and is also representative of its behaviour around the MW. No model is able to match all the observational constraints existing for the Sgr stream \citep{fardal19}. The observed difference in the orbital plane of Sgr and Sgr~II could suggest that the behaviour of the distant Sgr stream wrap that Sgr~II would be associated to is not perfectly described by the \citet{law_majewski10} simulation.
\end{itemize}

If either the second or the third scenario is the valid one, it would mean that Sgr~II is a new, exciting example of satellite of a satellite. Similarly to the Magellanic Clouds, the Sgr dSph would then have brought its own cohort of satellites that have now been deposited in the MW halo. Moreover, it would also bring some precious insights on the orbit of the Sgr stream in regions where it is poorly constrained.

\newpage

\section*{Acknowledgments}
We gratefully thank the CFHT staff for performing the observations in queue mode, for their reactivity in adapting the schedule, and for answering our questions during the data-reduction process. We thank Nina Hernitschek for granting us access to the catalogue of Pan-STARRS variability catalogue. 

ES, KY, and AA gratefully acknowledge funding by the Emmy Noether program from the Deutsche Forschungsgemeinschaft (DFG). This work has been published under the framework of the IdEx Unistra and benefits from a funding from the state managed by the French National Research Agency as part of the investments for the future program. NFM, RI, and NL gratefully acknowledge support from the French National Research Agency (ANR) funded project ``Pristine'' (ANR-18-CE31-0017) along with funding from CNRS/INSU through the Programme National Galaxies et Cosmologie and through the CNRS grant PICS07708. The authors thank the International Space Science Institute, Berne, Switzerland for providing financial support and meeting facilities to the international team ``Pristine''. JIGH acknowledges financial support from the Spanish Ministry project MINECO AYA2017-86389-P, and from the Spanish MINECO under the 2013 Ram\'on y Cajal program MINECO RYC-2013-14875.
 
Based on observations obtained at the Canada-France-Hawaii Telescope (CFHT) which is operated by the National Research Council of Canada, the Institut National des Sciences de l'Univers of the Centre National de la Recherche Scientifique of France, and the University of Hawaii.

Some of the data presented herein were obtained at the W. M. Keck Observatory, which is operated as a scientific partnership among the California Institute of Technology, the University of California and the National Aeronautics and Space Administration. The Observatory was made possible by the generous financial support of the W. M. Keck Foundation. Furthermore, the authors wish to recognize and acknowledge the very significant cultural role and reverence that the summit of Maunakea has always had within the indigenous Hawaiian community.  We are most fortunate to have the opportunity to conduct observations from this mountain.

The Pan-STARRS1 Surveys (PS1) have been made possible through contributions of the Institute for Astronomy, the University of Hawaii, the Pan-STARRS Project Office, the Max-Planck Society and its participating institutes, the Max Planck Institute for Astronomy, Heidelberg and the Max Planck Institute for Extraterrestrial Physics, Garching, The Johns Hopkins University, Durham University, the University of Edinburgh, Queen's University Belfast, the Harvard-Smithsonian Center for Astrophysics, the Las Cumbres Observatory Global Telescope Network Incorporated, the National Central University of Taiwan, the Space Telescope Science Institute, the National Aeronautics and Space Administration under Grant No. NNX08AR22G issued through the Planetary Science Division of the NASA Science Mission Directorate, the National Science Foundation under Grant No. AST-1238877, the University of Maryland, and Eotvos Lorand University (ELTE).

This work has made use of data from the European Space Agency (ESA)
mission {\it Gaia} (\url{https://www.cosmos.esa.int/gaia}), processed by
the {\it Gaia} Data Processing and Analysis Consortium (DPAC,
\url{https://www.cosmos.esa.int/web/gaia/dpac/consortium}). Funding
for the DPAC has been provided by national institutions, in particular
the institutions participating in the {\it Gaia} Multilateral Agreement.

\clearpage

\newpage

\begin{table*}

\caption{Properties of our spectroscopic sample. The Pristine metallicity of a given star is indicated only if [Fe/H]$_\mathrm{CaHK} < -1.0$. The individual spectroscopic metallicity is reported for stars with S/N $>= 12$ and $g_0 > 20.5$ only. Stars with $P_{mem} > 0.8$ are systematically considered as members. Potential horizontal branch stars of Sgr~II are marked as ``HB'' as the spectroscopic pipeline extracting the velocities is less reliable for those stars. Since our CMD fitting procedure described in section 4.1 does not account for the horizontal branch, their membership probability is not meaningful. Potential binary stars (as defined in section 2.2) are marked as ``B''. The systematic threshold $\delta_{\mathrm{thr}}$ is not included in the velocity uncertainties presented in this table.
\label{tbl-2}}

\setlength{\tabcolsep}{2.5pt}
\renewcommand{\arraystretch}{0.3}
\begin{sideways}
\begin{tabular}{cccccccccccccc}
\hline
RA (deg) & DEC (deg) & $g_0$ & $i_0$ & $CaHK_0$ & $v_{r} (\kms)$ & $\mu_{\alpha}^{*}$ (mas.yr$^{-1}$) & $\mu_{\delta}$ (mas.yr$^{-1}$) &  S/N & [Fe/H]$_{\mathrm{CaHK}}$ & [Fe/H]$_\mathrm{spectro}$ & $P_{mem}$ & Member\\
\hline

298.20599167 & $-$21.98790000 & 18.53 $\pm$ 0.01 & 17.99 $\pm$ 0.01 & 19.44 $\pm$ 0.01 & 16.6 $\pm$ 1.3 & 2.949 $\pm$ 0.506 & -6.861 $\pm$ 0.274 & 29.2 & --- & --- & 0.00 &  N  \\ \\ 
298.13157500 & $-$21.98582778 & 19.00 $\pm$ 0.01 & 18.54 $\pm$ 0.01 & 19.77 $\pm$ 0.01 & 27.5 $\pm$ 1.5 & -5.194 $\pm$ 0.738 & -4.88 $\pm$ 0.384 & 22.5 & --- & --- & 0.00 &  N  \\ \\ 
298.13137917 & $-$21.98273333 & 19.16 $\pm$ 0.01 & 18.50 $\pm$ 0.01 & 20.13 $\pm$ 0.01 & -39.4 $\pm$ 2.2 & 0.336 $\pm$ 0.69 & -7.317 $\pm$ 0.38 & 25.8 & $-$1.39 $\pm$ 0.10 & --- & 0.00 &  N  \\ \\ 
298.15507917 & $-$21.98054167 & 19.34 $\pm$ 0.01 & 18.66 $\pm$ 0.01 & 20.52 $\pm$ 0.02 & 12.5 $\pm$ 1.7 & 1.159 $\pm$ 0.889 & -8.058 $\pm$ 0.507 & 21.2 & --- & --- & 0.00 &  N  \\ \\ 
298.15097500 & $-$21.95283333 & 19.46 $\pm$ 0.01 & 18.90 $\pm$ 0.01 & 20.39 $\pm$ 0.02 & 140.9 $\pm$ 1.7 & -6.681 $\pm$ 1.002 & 0.739 $\pm$ 0.503 & 21.5 & $-$1.02 $\pm$ 0.15 & --- & 0.00 &  N  \\ \\ 
298.19072083 & $-$21.96758056 & 19.88 $\pm$ 0.01 & 19.54 $\pm$ 0.01 & 20.19 $\pm$ 0.01 & -118.1 $\pm$ 6.2 & -0.867 $\pm$ 1.551 & 1.216 $\pm$ 0.81 & 14.4 & $-$3.61 $\pm$ 0.17 & -2.85 $\pm$ 0.23 & 0.00 &  N  \\ \\ 
298.18688333 & $-$21.97713889 & 20.43 $\pm$ 0.01 & 19.87 $\pm$ 0.01 & 21.41 $\pm$ 0.03 & -107.3 $\pm$ 2.8 & -5.917 $\pm$ 2.251 & -10.725 $\pm$ 1.175 & 12.6 & --- & -1.26 $\pm$ 0.12 & 0.00 &  N  \\ \\ 
298.18320417 & $-$21.96361944 & 21.28 $\pm$ 0.01 & 20.73 $\pm$ 0.01 & 22.22 $\pm$ 0.06 & -15.4 $\pm$ 3.5 & --- & --- & 5.6 & --- & --- & 0.30 &  N  \\ \\ 
298.13665417 & $-$21.97423889 & 21.41 $\pm$ 0.01 & 20.90 $\pm$ 0.01 & 22.18 $\pm$ 0.06 & -7.8 $\pm$ 7.4 & --- & --- & 4.7 & $-$1.40 $\pm$ 0.30 & --- & 0.04 &  N  \\ \\ 
298.19440000 & $-$21.99070556 & 21.60 $\pm$ 0.01 & 21.22 $\pm$ 0.02 & 22.17 $\pm$ 0.06 & -67.3 $\pm$ 3.8 & --- & --- & 3.9 & $-$1.45 $\pm$ 0.35 & --- & 0.00 &  N  \\ \\ 
298.19622500 & $-$21.99368889 & 22.40 $\pm$ 0.02 & 21.66 $\pm$ 0.02 & 23.62 $\pm$ 0.18 & -2.0 $\pm$ 13.4 & --- & --- & 2.8 & --- & --- & 0.00 &  N  \\ \\ 
298.16959583 & $-$22.17463056 & 17.71 $\pm$ 0.01 & 16.94 $\pm$ 0.01 & 19.06 $\pm$ 0.01 & 89.7 $\pm$ 1.1 & -3.409 $\pm$ 0.228 & -5.191 $\pm$ 0.172 & 31.4 & --- & --- & 0.00 &  N  \\ \\ 
298.18001250 & $-$22.07175000 & 17.90 $\pm$ 0.01 & 17.05 $\pm$ 0.01 & 19.32 $\pm$ 0.01 & 0.2 $\pm$ 0.9 & 1.896 $\pm$ 0.225 & -7.431 $\pm$ 0.133 & 33.5 & --- & --- & 0.00 &  N  \\ \\ 
298.16810000 & $-$22.18830000 & 18.27 $\pm$ 0.01 & 17.65 $\pm$ 0.01 & 19.27 $\pm$ 0.01 & 105.1 $\pm$ 1.3 & -3.679 $\pm$ 0.287 & -3.852 $\pm$ 0.18 & 37.2 & $-$1.01 $\pm$ 0.11 & --- & 0.00 &  N  \\ \\ 
298.19661667 & $-$22.14635278 & 18.73 $\pm$ 0.01 & 18.31 $\pm$ 0.01 & 19.38 $\pm$ 0.01 & 29.2 $\pm$ 1.8 & -0.708 $\pm$ 0.441 & -4.393 $\pm$ 0.262 & 21.1 & $-$1.30 $\pm$ 0.13 & --- & 0.00 &  N  \\ \\ 
298.17483750 & $-$22.16606389 & 18.82 $\pm$ 0.01 & 18.33 $\pm$ 0.01 & 19.65 $\pm$ 0.01 & -54.0 $\pm$ 1.3 & 1.647 $\pm$ 0.507 & -3.121 $\pm$ 0.37 & 27.6 & --- & --- & 0.00 &  N  \\ \\ 
298.19189583 & $-$22.19686944 & 18.90 $\pm$ 0.01 & 18.34 $\pm$ 0.01 & 19.79 $\pm$ 0.01 & 65.8 $\pm$ 4.0 & -1.137 $\pm$ 0.558 & -3.422 $\pm$ 0.38 & 36.6 & $-$1.15 $\pm$ 0.12 & --- & 0.00 &  N  \\ \\ 
298.19827500 & $-$22.14498333 & 18.94 $\pm$ 0.01 & 18.35 $\pm$ 0.01 & 19.77 $\pm$ 0.01 & 89.3 $\pm$ 1.3 & -1.161 $\pm$ 0.478 & -7.535 $\pm$ 0.281 & 25.3 & $-$1.54 $\pm$ 0.13 & --- & 0.00 &  N  \\ \\ 
298.16145833 & $-$22.08265556 & 18.98 $\pm$ 0.01 & 18.20 $\pm$ 0.01 & 19.83 $\pm$ 0.01 & -182.8 $\pm$ 0.9 & -0.27 $\pm$ 0.46 & -0.825 $\pm$ 0.269 & 26.3 & $-$2.62 $\pm$ 0.13 & -2.27 $\pm$ 0.04 & 0.98 &  Y  \\ \\ 
298.16120000 & $-$22.00828611 & 19.04 $\pm$ 0.01 & 18.41 $\pm$ 0.01 & 20.13 $\pm$ 0.01 & -85.1 $\pm$ 1.4 & 3.636 $\pm$ 0.626 & -8.935 $\pm$ 0.343 & 26.7 & --- & --- & 0.00 &  N  \\ \\ 
298.15403750 & $-$22.11108333 & 19.17 $\pm$ 0.01 & 18.71 $\pm$ 0.01 & 19.92 $\pm$ 0.01 & 51.4 $\pm$ 1.9 & 1.085 $\pm$ 0.593 & -4.926 $\pm$ 0.335 & 17.5 & $-$1.02 $\pm$ 0.15 & --- & 0.00 &  N  \\ \\ 
298.15852917 & $-$22.05846944 & 19.42 $\pm$ 0.01 & 19.59 $\pm$ 0.01 & 19.67 $\pm$ 0.01 & -167.9 $\pm$ 9.5 & 0.865 $\pm$ 1.014 & -0.73 $\pm$ 0.571 & 5.0 & --- & -0.43 $\pm$ 0.29 & 0.00 &  HB  \\ \\ 
298.19296667 & $-$22.02218889 & 19.44 $\pm$ 0.01 & 18.93 $\pm$ 0.01 & 20.21 $\pm$ 0.01 & 122.6 $\pm$ 1.8 & -0.935 $\pm$ 0.985 & -1.78 $\pm$ 0.535 & 17.8 & $-$1.39 $\pm$ 0.13 & --- & 0.00 &  N  \\ \\ 
298.15097500 & $-$22.07938611 & 19.56 $\pm$ 0.01 & 19.17 $\pm$ 0.01 & 19.94 $\pm$ 0.01 & -174.0 $\pm$ 2.1 & -1.205 $\pm$ 0.805 & -0.816 $\pm$ 0.463 & 16.9 & $-$3.16 $\pm$ 0.29 & -2.57 $\pm$ 0.14 & 0.00 &  N  \\ \\ 
298.17763750 & $-$22.04601111 & 19.62 $\pm$ 0.01 & 19.83 $\pm$ 0.01 & 19.63 $\pm$ 0.01 & -135.3 $\pm$ 4.2 & -4.001 $\pm$ 1.206 & -1.43 $\pm$ 0.651 & 11.1 & --- & -0.92 $\pm$ 0.28 & 0.00 &  HB  \\ \\ 
298.17252500 & $-$22.07410833 & 19.66 $\pm$ 0.01 & 19.90 $\pm$ 0.01 & 19.67 $\pm$ 0.01 & -76.1 $\pm$ 10.3 & -0.958 $\pm$ 1.284 & -2.143 $\pm$ 0.663 & 11.7 & --- & -0.05 $\pm$ 0.23 & 0.00 &  HB  \\ \\ 
298.12763750 & $-$22.17288611 & 19.75 $\pm$ 0.01 & 19.37 $\pm$ 0.01 & 20.20 $\pm$ 0.01 & -15.5 $\pm$ 2.6 & 2.824 $\pm$ 0.831 & -6.75 $\pm$ 0.504 & 15.0 & $-$2.38 $\pm$ 0.29 & -1.77 $\pm$ 0.16 & 0.00 &  N  \\ \\ 
298.20524167 & $-$22.02750556 & 19.83 $\pm$ 0.01 & 19.01 $\pm$ 0.01 & 21.05 $\pm$ 0.02 & 201.1 $\pm$ 3.4 & -3.185 $\pm$ 1.126 & -15.463 $\pm$ 0.607 & 17.8 & $-$1.27 $\pm$ 0.13 & --- & 0.00 &  B  \\ \\ 
298.18212917 & $-$22.05709444 & 19.85 $\pm$ 0.01 & 19.04 $\pm$ 0.01 & 21.08 $\pm$ 0.03 & 28.4 $\pm$ 1.8 & -0.533 $\pm$ 1.135 & -6.337 $\pm$ 0.594 & 14.9 & $-$1.24 $\pm$ 0.12 & --- & 0.03 &  N  \\ \\ 
298.14820833 & $-$22.00248056 & 19.89 $\pm$ 0.01 & 19.52 $\pm$ 0.01 & 20.39 $\pm$ 0.02 & 163.6 $\pm$ 1.9 & 0.014 $\pm$ 1.404 & -4.599 $\pm$ 0.742 & 14.3 & $-$1.82 $\pm$ 0.23 & -2.5 $\pm$ 0.12 & 0.00 &  N  \\ \\ 
298.16396667 & $-$22.06349722 & 19.94 $\pm$ 0.01 & 19.53 $\pm$ 0.01 & 20.46 $\pm$ 0.02 & 41.4 $\pm$ 2.0 & 3.398 $\pm$ 1.562 & -8.363 $\pm$ 0.947 & 14.0 & $-$2.04 $\pm$ 0.25 & -2.01 $\pm$ 0.11 & 0.00 &  N  \\ \\ 
298.16217500 & $-$22.05441111 & 19.96 $\pm$ 0.01 & 19.24 $\pm$ 0.01 & 20.72 $\pm$ 0.02 & -176.0 $\pm$ 1.5 & -1.281 $\pm$ 1.022 & -1.531 $\pm$ 0.539 & 15.8 & $-$2.65 $\pm$ 0.16 & -2.24 $\pm$ 0.08 & 0.99 &  Y  \\ \\ 
298.14762500 & $-$22.18983611 & 20.06 $\pm$ 0.01 & 19.67 $\pm$ 0.01 & 20.63 $\pm$ 0.02 & -227.3 $\pm$ 3.3 & -1.419 $\pm$ 1.191 & -5.42 $\pm$ 0.742 & 11.3 & $-$1.57 $\pm$ 0.18 & --- & 0.00 &  N  \\ \\ 
298.19400833 & $-$22.08638333 & 20.38 $\pm$ 0.01 & 19.70 $\pm$ 0.01 & 21.09 $\pm$ 0.03 & -174.9 $\pm$ 2.3 & -0.305 $\pm$ 1.333 & 0.776 $\pm$ 0.766 & 13.2 & $-$2.59 $\pm$ 0.23 & -2.09 $\pm$ 0.11 & 0.99 &  Y  \\ \\ 
298.19723333 & $-$22.12452500 & 20.51 $\pm$ 0.01 & 19.94 $\pm$ 0.01 & 21.45 $\pm$ 0.03 & -323.9 $\pm$ 2.6 & -10.391 $\pm$ 1.957 & -3.304 $\pm$ 1.107 & 12.0 & --- & --- & 0.01 &  N  \\ \\ 
298.14902500 & $-$22.02358611 & 20.91 $\pm$ 0.01 & 20.49 $\pm$ 0.01 & 21.45 $\pm$ 0.03 & -71.1 $\pm$ 3.6 & --- & --- & 7.2 & $-$2.05 $\pm$ 0.31 & --- & 0.00 &  N  \\ \\ 
298.18112083 & $-$22.06077222 & 20.93 $\pm$ 0.01 & 20.29 $\pm$ 0.01 & 21.60 $\pm$ 0.04 & -179.5 $\pm$ 2.4 & --- & --- & 9.7 & $-$2.54 $\pm$ 0.28 & --- & 1.00 &  Y  \\ \\ 
298.17317500 & $-$22.11583611 & 21.10 $\pm$ 0.01 & 20.51 $\pm$ 0.01 & 21.83 $\pm$ 0.04 & -173.9 $\pm$ 3.8 & --- & --- & 6.2 & $-$2.06 $\pm$ 0.22 & --- & 0.97 &  B  \\ \\ 
298.14954583 & $-$22.10703056 & 21.24 $\pm$ 0.01 & 20.62 $\pm$ 0.01 & 21.89 $\pm$ 0.05 & -177.4 $\pm$ 3.7 & --- & --- & 6.2 & $-$2.69 $\pm$ 0.45 & --- & 0.99 &  Y  \\ \\ 
298.18245417 & $-$22.10563889 & 21.27 $\pm$ 0.01 & 20.67 $\pm$ 0.01 & 21.89 $\pm$ 0.05 & -176.3 $\pm$ 6.0 & --- & --- & 5.6 & $-$2.67 $\pm$ 0.36 & --- & 0.99 &  Y  \\ \\ 
298.15165833 & $-$22.14928333 & 21.54 $\pm$ 0.01 & 21.13 $\pm$ 0.02 & 22.07 $\pm$ 0.05 & 10.6 $\pm$ 10.6 & --- & --- & 4.6 & $-$1.82 $\pm$ 0.51 & --- & 0.00 &  N  \\ \\ 
298.16334583 & $-$22.14321667 & 21.60 $\pm$ 0.01 & 21.16 $\pm$ 0.02 & 22.27 $\pm$ 0.06 & 71.0 $\pm$ 4.9 & --- & --- & 4.0 & $-$1.31 $\pm$ 0.36 & --- & 0.00 &  N  \\ \\ 
298.16188333 & $-$22.05321389 & 21.56 $\pm$ 0.01 & 20.99 $\pm$ 0.01 & 22.31 $\pm$ 0.07 & -177.5 $\pm$ 7.8 & --- & --- & 5.7 & $-$1.81 $\pm$ 0.37 & --- & 1.00 &  Y  \\ \\ 
298.18229167 & $-$22.04275278 & 21.61 $\pm$ 0.01 & 20.99 $\pm$ 0.01 & 22.22 $\pm$ 0.06 & -177.1 $\pm$ 3.9 & --- & --- & 5.5 & $-$2.96 $\pm$ 0.47 & --- & 0.99 &  Y  \\ \\ 
298.19592917 & $-$22.13295000 & 21.65 $\pm$ 0.01 & 21.03 $\pm$ 0.01 & 22.88 $\pm$ 0.10 & 22.0 $\pm$ 6.6 & --- & --- & -0.3 & --- & --- & 0.90 &  N  \\ \\ 
298.19628750 & $-$22.18246389 & 21.79 $\pm$ 0.01 & 21.30 $\pm$ 0.02 & 22.57 $\pm$ 0.08 & -285.7 $\pm$ 12.8 & --- & --- & 4.2 & $-$1.33 $\pm$ 0.37 & --- & 0.02 &  N  \\ \\ 
298.13948750 & $-$22.17620000 & 21.81 $\pm$ 0.01 & 21.41 $\pm$ 0.02 & 22.37 $\pm$ 0.07 & 174.5 $\pm$ 11.6 & --- & --- & 2.6 & $-$1.49 $\pm$ 0.43 & --- & 0.00 &  N  \\ \\ 
298.15003333 & $-$22.01741667 & 21.77 $\pm$ 0.01 & 21.01 $\pm$ 0.01 & 22.59 $\pm$ 0.08 & -276.5 $\pm$ 3.6 & --- & --- & 4.5 & $-$2.68 $\pm$ 0.56 & --- & 0.00 &  N  \\ \\ 
298.14941250 & $-$22.17936667 & 21.82 $\pm$ 0.01 & 21.21 $\pm$ 0.02 & 22.92 $\pm$ 0.11 & -59.4 $\pm$ 5.6 & --- & --- & 4.1 & --- & --- & 0.57 &  N  \\ \\ 
298.16383333 & $-$22.18616667 & 21.90 $\pm$ 0.01 & 21.11 $\pm$ 0.02 & 23.39 $\pm$ 0.16 & -11.5 $\pm$ 5.2 & --- & --- & 3.3 & --- & --- & 0.00 &  N  \\ \\ 
298.16461667 & $-$22.09165278 & 21.93 $\pm$ 0.01 & 21.41 $\pm$ 0.02 & 22.58 $\pm$ 0.08 & -172.1 $\pm$ 4.5 & --- & --- & 3.7 & $-$2.05 $\pm$ 0.41 & --- & 1.00 &  Y  \\ \\ 
298.19036667 & $-$22.09013056 & 22.25 $\pm$ 0.02 & 21.75 $\pm$ 0.02 & 22.90 $\pm$ 0.11 & -178.8 $\pm$ 4.4 & --- & --- & 2.6 & $-$1.83 $\pm$ 0.69 & --- & 1.00 &  Y  \\ \\

\end{tabular}
\end{sideways}
\end{table*}

\newpage

\newpage

\begin{table*}
\renewcommand\thetable{2} 
\caption{Properties of our spectroscopic sample - Part 2
\label{tbl-2}}

\setlength{\tabcolsep}{2.5pt}
\renewcommand{\arraystretch}{0.3}
\begin{sideways}
\begin{tabular}{cccccccccccccc}
\hline
RA (deg) & DEC (deg) & $g_0$ & $i_0$ & $CaHK_0$ & $v_{r} (\kms)$ & $\mu_{\alpha}^{*}$ (mas.yr$^{-1}$) & $\mu_{\delta}$ (mas.yr$^{-1}$) &  S/N & [Fe/H]$_{\mathrm{CaHK}}$ & [Fe/H]$_\mathrm{spectro}$ & $P_{mem}$ & Member\\
\hline

298.17171250 & $-$22.12262222 & 22.32 $\pm$ 0.02 & 21.53 $\pm$ 0.02 & 23.38 $\pm$ 0.16 & -109.2 $\pm$ 9.5 & --- & --- & 3.4 & $-$1.74 $\pm$ 0.52 & --- & 0.00 &  N  \\ \\ 
298.12890833 & $-$22.15738333 & 22.55 $\pm$ 0.02 & 21.04 $\pm$ 0.01 & 22.81 $\pm$ 0.10 & 31.0 $\pm$ 6.0 & --- & --- & 4.8 & --- & --- & 0.00 &  N  \\ \\ 
298.19759167 & $-$22.11256944 & 22.63 $\pm$ 0.02 & 22.04 $\pm$ 0.03 & $-$0.420 $\pm$ 0.00 & 106.2 $\pm$ 4.9 & --- & --- & 2.2 & --- & --- & 0.00 &  N  \\ \\ 
298.18128333 & $-$22.11434167 & 22.70 $\pm$ 0.02 & 21.87 $\pm$ 0.02 & $-$0.410 $\pm$ 0.00 & -15.4 $\pm$ 10.0 & --- & --- & 2.7 & --- & --- & 0.00 &  N  \\ \\ 
298.13974583 & $-$22.02536944 & 22.77 $\pm$ 0.02 & 22.02 $\pm$ 0.03 & 23.42 $\pm$ 0.16 & 642.6 $\pm$ 13.9 & --- & --- & 2.4 & --- & --- & 0.00 &  N  \\ \\ 
298.16238403 & $-$22.07748222 & 17.50 $\pm$ 0.01 & 16.37 $\pm$ 0.01 & 18.97 $\pm$ 0.01 & -170.4 $\pm$ 0.7 & -0.548 $\pm$ 0.158 & -0.878 $\pm$ 0.087 & 50.6 & --- & -2.09 $\pm$ 0.04 & 1.00 &  Y  \\ \\ 
298.16424561 & $-$22.16802978 & 17.47 $\pm$ 0.01 & 16.68 $\pm$ 0.01 & 18.82 $\pm$ 0.01 & -82.8 $\pm$ 1.1 & 14.405 $\pm$ 0.181 & -4.936 $\pm$ 0.108 & 38.8 & --- & -1.39 $\pm$ 0.06 & 0.00 &  N  \\ \\ 
298.29534583 & $-$22.07591389 & 18.37 $\pm$ 0.01 & 17.55 $\pm$ 0.01 & 19.82 $\pm$ 0.01 & 109.3 $\pm$ 2.6 & -4.564 $\pm$ 0.37 & -3.604 $\pm$ 0.22 & 32.2 & --- & --- & 0.00 &  N  \\ \\ 
298.26279167 & $-$22.10196944 & 18.50 $\pm$ 0.01 & 17.87 $\pm$ 0.01 & 19.58 $\pm$ 0.01 & 66.3 $\pm$ 5.3 & -3.609 $\pm$ 0.38 & -9.306 $\pm$ 0.209 & 15.3 & --- & --- & 0.00 &  N  \\ \\ 
298.26702500 & $-$22.08325833 & 19.00 $\pm$ 0.01 & 18.54 $\pm$ 0.01 & 19.80 $\pm$ 0.01 & 11.9 $\pm$ 1.6 & 3.988 $\pm$ 0.662 & -6.709 $\pm$ 0.369 & 24.3 & --- & --- & 0.00 &  N  \\ \\ 
298.31595000 & $-$22.06791667 & 19.21 $\pm$ 0.01 & 18.70 $\pm$ 0.01 & 19.84 $\pm$ 0.01 & -85.3 $\pm$ 2.6 & 2.98 $\pm$ 0.722 & -7.838 $\pm$ 0.438 & 26.2 & $-$2.01 $\pm$ 0.15 & -1.77 $\pm$ 0.09 & 0.00 &  N  \\ \\ 
298.26878333 & $-$22.02816111 & 19.22 $\pm$ 0.01 & 18.54 $\pm$ 0.01 & 20.39 $\pm$ 0.02 & 42.9 $\pm$ 1.8 & -5.562 $\pm$ 0.769 & -9.669 $\pm$ 0.406 & 24.0 & --- & --- & 0.00 &  N  \\ \\ 
298.28164167 & $-$22.08365556 & 19.48 $\pm$ 0.01 & 18.95 $\pm$ 0.01 & 20.34 $\pm$ 0.02 & 31.1 $\pm$ 2.3 & -1.95 $\pm$ 0.912 & -7.823 $\pm$ 0.498 & 22.5 & $-$1.01 $\pm$ 0.14 & --- & 0.00 &  N  \\ \\ 
298.30703333 & $-$22.09047222 & 19.48 $\pm$ 0.01 & 18.62 $\pm$ 0.01 & 20.98 $\pm$ 0.02 & 33.2 $\pm$ 3.6 & -4.498 $\pm$ 0.838 & 0.209 $\pm$ 0.47 & 30.8 & --- & --- & 0.00 &  N  \\ \\ 
298.28935417 & $-$22.06904444 & 19.59 $\pm$ 0.01 & 19.18 $\pm$ 0.01 & 20.19 $\pm$ 0.01 & 17.8 $\pm$ 1.7 & 1.286 $\pm$ 1.111 & -5.091 $\pm$ 0.609 & 22.4 & $-$1.52 $\pm$ 0.18 & --- & 0.00 &  N  \\ \\ 
298.29140833 & $-$22.07521389 & 19.68 $\pm$ 0.01 & 19.31 $\pm$ 0.01 & 20.18 $\pm$ 0.01 & -40.9 $\pm$ 2.0 & -2.302 $\pm$ 1.145 & -11.191 $\pm$ 0.665 & 22.0 & $-$1.82 $\pm$ 0.22 & -2.19 $\pm$ 0.11 & 0.00 &  N  \\ \\ 
298.26110000 & $-$22.03198333 & 19.79 $\pm$ 0.01 & 19.16 $\pm$ 0.01 & 20.86 $\pm$ 0.02 & -108.8 $\pm$ 2.2 & 4.747 $\pm$ 1.23 & -6.477 $\pm$ 0.661 & 21.6 & --- & --- & 0.00 &  N  \\ \\ 
298.27997917 & $-$22.07672222 & 20.40 $\pm$ 0.01 & 19.75 $\pm$ 0.01 & 21.44 $\pm$ 0.03 & -111.5 $\pm$ 3.2 & -5.707 $\pm$ 2.05 & -4.735 $\pm$ 1.075 & 15.6 & --- & --- & 0.49 &  N  \\ \\ 
298.25045417 & $-$22.08092778 & 20.39 $\pm$ 0.01 & 19.57 $\pm$ 0.01 & 21.80 $\pm$ 0.04 & 73.8 $\pm$ 2.4 & -4.985 $\pm$ 1.811 & -11.227 $\pm$ 0.907 & 18.1 & --- & --- & 0.00 &  N  \\ \\ 
298.25660833 & $-$22.04040278 & 20.73 $\pm$ 0.01 & 20.19 $\pm$ 0.01 & 21.42 $\pm$ 0.03 & -33.3 $\pm$ 4.3 & 5.009 $\pm$ 3.983 & -5.137 $\pm$ 2.092 & 10.1 & $-$1.93 $\pm$ 0.20 & --- & 0.00 &  N  \\ \\ 
298.23870417 & $-$22.10072500 & 21.06 $\pm$ 0.01 & 20.60 $\pm$ 0.01 & 21.83 $\pm$ 0.04 & -85.8 $\pm$ 5.1 & --- & --- & 8.4 & --- & --- & 0.00 &  N  \\ \\ 
298.25472083 & $-$22.05189444 & 21.49 $\pm$ 0.01 & 20.99 $\pm$ 0.01 & 22.09 $\pm$ 0.05 & -103.4 $\pm$ 5.5 & --- & --- & 6.5 & $-$2.30 $\pm$ 0.40 & --- & 0.04 &  N  \\ \\ 
298.31194583 & $-$22.06705278 & 21.50 $\pm$ 0.01 & 20.63 $\pm$ 0.01 & 23.09 $\pm$ 0.12 & 125.8 $\pm$ 6.7 & --- & --- & 8.5 & --- & --- & 0.00 &  N  \\ \\ 
298.29775417 & $-$22.10044444 & 21.77 $\pm$ 0.01 & 21.15 $\pm$ 0.02 & 22.69 $\pm$ 0.08 & 67.2 $\pm$ 3.8 & --- & --- & 5.8 & $-$1.36 $\pm$ 0.34 & --- & 0.35 &  N  \\ \\ 
298.27070417 & $-$22.09640278 & 22.06 $\pm$ 0.01 & 21.22 $\pm$ 0.02 & 23.49 $\pm$ 0.16 & -64.3 $\pm$ 7.1 & --- & --- & 5.7 & --- & --- & 0.00 &  N  \\ \\ 
298.28685000 & $-$22.04170000 & 22.40 $\pm$ 0.02 & 21.62 $\pm$ 0.02 & $-$0.420 $\pm$ 0.00 & 217.5 $\pm$ 5.7 & --- & --- & 3.7 & --- & --- & 0.00 &  N  \\ \\ 
298.11754583 & $-$22.07703889 & 18.09 $\pm$ 0.01 & 17.49 $\pm$ 0.01 & 19.07 $\pm$ 0.01 & 21.4 $\pm$ 1.5 & -2.317 $\pm$ 0.273 & -2.956 $\pm$ 0.152 & 28.9 & --- & --- & 0.00 &  N  \\ \\ 
298.21458333 & $-$22.09340000 & 18.59 $\pm$ 0.01 & 18.06 $\pm$ 0.01 & 19.63 $\pm$ 0.01 & 9.5 $\pm$ 2.5 & -1.798 $\pm$ 0.406 & -3.73 $\pm$ 0.234 & 25.2 & --- & --- & 0.00 &  N  \\ \\ 
298.15560000 & $-$22.05777778 & 19.22 $\pm$ 0.01 & 18.91 $\pm$ 0.01 & 19.58 $\pm$ 0.01 & 13.6 $\pm$ 2.3 & -1.788 $\pm$ 0.74 & -5.022 $\pm$ 0.38 & 20.8 & $-$2.71 $\pm$ 0.29 & -2.81 $\pm$ 0.17 & 0.00 &  N  \\ \\ 
298.12512917 & $-$22.04615556 & 19.24 $\pm$ 0.01 & 18.87 $\pm$ 0.01 & 19.63 $\pm$ 0.01 & -179.3 $\pm$ 4.2 & 9.328 $\pm$ 0.675 & -6.482 $\pm$ 0.376 & 24.2 & $-$3.08 $\pm$ 0.27 & -3.7 $\pm$ 0.33 & 0.00 &  N  \\ \\ 
298.16298750 & $-$22.08243889 & 19.38 $\pm$ 0.01 & 18.61 $\pm$ 0.01 & 20.18 $\pm$ 0.01 & -176.0 $\pm$ 1.3 & -0.547 $\pm$ 0.608 & -0.29 $\pm$ 0.338 & 26.6 & $-$2.73 $\pm$ 0.16 & -2.42 $\pm$ 0.09 & 0.99 &  Y  \\ \\ 
298.22307917 & $-$22.03975556 & 19.52 $\pm$ 0.01 & 18.78 $\pm$ 0.01 & 20.76 $\pm$ 0.02 & 60.9 $\pm$ 1.9 & -5.413 $\pm$ 0.935 & -5.874 $\pm$ 0.519 & 21.5 & --- & --- & 0.92 &  N  \\ \\ 
298.18333333 & $-$22.07022778 & 19.79 $\pm$ 0.01 & 19.33 $\pm$ 0.01 & 20.41 $\pm$ 0.02 & -110.3 $\pm$ 2.0 & 7.892 $\pm$ 0.932 & 4.142 $\pm$ 0.541 & 18.4 & $-$1.79 $\pm$ 0.19 & -1.86 $\pm$ 0.09 & 0.00 &  N  \\ \\ 
298.17679167 & $-$22.05593889 & 19.98 $\pm$ 0.01 & 19.16 $\pm$ 0.01 & 21.27 $\pm$ 0.03 & -63.7 $\pm$ 2.4 & 2.017 $\pm$ 0.904 & -9.667 $\pm$ 0.501 & 21.1 & --- & --- & 0.00 &  N  \\ \\ 
298.16640833 & $-$22.06769444 & 20.21 $\pm$ 0.01 & 19.51 $\pm$ 0.01 & 20.95 $\pm$ 0.02 & -176.0 $\pm$ 2.6 & 0.954 $\pm$ 1.292 & 0.563 $\pm$ 0.684 & 17.2 & $-$2.59 $\pm$ 0.14 & -2.31 $\pm$ 0.11 & 1.00 &  Y  \\ \\ 
298.17500000 & $-$22.05421389 & 20.70 $\pm$ 0.01 & 20.02 $\pm$ 0.01 & 21.41 $\pm$ 0.03 & -174.6 $\pm$ 1.7 & 1.235 $\pm$ 3.375 & -2.706 $\pm$ 1.621 & 13.1 & $-$2.66 $\pm$ 0.30 & --- & 0.99 &  Y  \\ \\ 
298.20856250 & $-$22.06173611 & 21.18 $\pm$ 0.01 & 20.65 $\pm$ 0.01 & 21.80 $\pm$ 0.04 & 107.6 $\pm$ 5.3 & --- & --- & 8.4 & $-$2.22 $\pm$ 0.29 & --- & 0.57 &  N  \\ \\ 
298.20445833 & $-$22.05373333 & 21.28 $\pm$ 0.01 & 20.55 $\pm$ 0.01 & 22.41 $\pm$ 0.07 & -164.6 $\pm$ 6.1 & --- & --- & 8.1 & $-$1.12 $\pm$ 0.25 & --- & 0.00 &  N  \\ \\ 
298.15244167 & $-$22.06164167 & 21.31 $\pm$ 0.01 & 20.72 $\pm$ 0.01 & 21.94 $\pm$ 0.05 & -177.4 $\pm$ 2.6 & --- & --- & 7.8 & $-$2.50 $\pm$ 0.41 & --- & 1.00 &  Y  \\ \\ 
298.19703750 & $-$22.07148333 & 21.65 $\pm$ 0.01 & 21.10 $\pm$ 0.02 & 22.28 $\pm$ 0.06 & -173.9 $\pm$ 6.0 & --- & --- & 6.6 & $-$2.45 $\pm$ 0.44 & --- & 1.00 &  Y  \\ \\ 
298.14088333 & $-$22.04198611 & 21.67 $\pm$ 0.01 & 21.08 $\pm$ 0.02 & 22.38 $\pm$ 0.07 & -184.7 $\pm$ 3.7 & --- & --- & 5.7 & $-$2.09 $\pm$ 0.44 & --- & 1.00 &  Y  \\ \\ 
298.11370417 & $-$22.07934444 & 21.75 $\pm$ 0.01 & 20.93 $\pm$ 0.01 & 23.21 $\pm$ 0.14 & -74.9 $\pm$ 5.7 & --- & --- & 6.9 & --- & --- & 0.00 &  N  \\ \\ 
298.13356250 & $-$22.10374444 & 21.98 $\pm$ 0.01 & 21.06 $\pm$ 0.01 & 23.45 $\pm$ 0.17 & -198.2 $\pm$ 10.4 & --- & --- & 5.2 & --- & --- & 0.00 &  N  \\ \\ 
298.20068333 & $-$22.05971667 & 22.05 $\pm$ 0.01 & 21.50 $\pm$ 0.02 & 22.79 $\pm$ 0.10 & -187.7 $\pm$ 5.8 & --- & --- & 3.8 & $-$1.73 $\pm$ 0.41 & --- & 1.00 &  Y  \\ \\ 
298.10810417 & $-$22.07218333 & 22.07 $\pm$ 0.01 & 21.51 $\pm$ 0.02 & 22.92 $\pm$ 0.11 & -193.6 $\pm$ 8.6 & --- & --- & 4.4 & $-$1.45 $\pm$ 0.60 & --- & 0.99 &  Y  \\ \\ 
298.19912083 & $-$22.08466111 & 22.15 $\pm$ 0.01 & 21.62 $\pm$ 0.02 & 22.90 $\pm$ 0.10 & -167.6 $\pm$ 11.5 & --- & --- & 3.7 & $-$1.63 $\pm$ 0.48 & --- & 1.00 &  Y  \\ \\ 
298.18915833 & $-$22.07193333 & 22.28 $\pm$ 0.02 & 21.83 $\pm$ 0.02 & 22.86 $\pm$ 0.10 & -17.5 $\pm$ 14.9 & --- & --- & 3.3 & $-$2.00 $\pm$ 0.73 & --- & 1.00 &  N  \\ \\ 
298.15436250 & $-$22.07077500 & 22.30 $\pm$ 0.02 & 21.79 $\pm$ 0.02 & 22.85 $\pm$ 0.10 & -171.8 $\pm$ 13.7 & --- & --- & 3.7 & $-$2.72 $\pm$ 0.81 & --- & 1.00 &  Y  \\ \\ 
298.14889167 & $-$22.04349167 & 22.36 $\pm$ 0.02 & 21.78 $\pm$ 0.02 & 23.45 $\pm$ 0.17 & -26.3 $\pm$ 13.5 & --- & --- & 3.1 & --- & --- & 0.99 &  N  \\ \\ 
298.08678333 & $-$22.10022500 & 18.37 $\pm$ 0.01 & 17.60 $\pm$ 0.01 & 19.64 $\pm$ 0.01 & 18.0 $\pm$ 1.5 & 0.656 $\pm$ 0.279 & -4.619 $\pm$ 0.163 & 28.0 & --- & --- & 0.00 &  N  \\ \\ 
298.06311667 & $-$22.05839722 & 18.67 $\pm$ 0.01 & 18.19 $\pm$ 0.01 & 19.47 $\pm$ 0.01 & 16.8 $\pm$ 1.6 & 5.475 $\pm$ 0.404 & -5.921 $\pm$ 0.228 & 28.9 & $-$1.10 $\pm$ 0.12 & --- & 0.00 &  N  \\ \\ 
298.08457083 & $-$22.04709722 & 19.01 $\pm$ 0.01 & 18.37 $\pm$ 0.01 & 20.16 $\pm$ 0.01 & 39.1 $\pm$ 1.9 & -2.449 $\pm$ 0.482 & -2.806 $\pm$ 0.268 & 27.5 & --- & --- & 0.00 &  N  \\ \\

\end{tabular}
\end{sideways}
\end{table*}

\newpage

\begin{table*}
\renewcommand\thetable{2} 
\caption{Properties of our spectroscopic sample - Part 3
\label{tbl-2}}
\setlength{\tabcolsep}{2.5pt}
\renewcommand{\arraystretch}{0.3}
\begin{sideways}
\begin{tabular}{cccccccccccccc}
\hline
RA (deg) & DEC (deg) & $g_0$ & $i_0$ & $CaHK_0$ & $v_{r} (\kms)$ & $\mu_{\alpha}^{*}$ (mas.yr$^{-1}$) & $\mu_{\delta}$ (mas.yr$^{-1}$) &  S/N & [Fe/H]$_{\mathrm{CaHK}}$ & [Fe/H]$_\mathrm{spectro}$ & $P_{mem}$ & Member\\
\hline

298.06152500 & $-$22.02591111 & 19.30 $\pm$ 0.01 & 18.59 $\pm$ 0.01 & 20.56 $\pm$ 0.02 & 48.4 $\pm$ 2.4 & -1.92 $\pm$ 0.559 & -3.82 $\pm$ 0.317 & 22.1 & --- & --- & 0.01 &  N  \\ \\ 
298.08085833 & $-$22.04420000 & 19.32 $\pm$ 0.01 & 18.79 $\pm$ 0.01 & 19.89 $\pm$ 0.01 & -385.3 $\pm$ 1.6 & -1.849 $\pm$ 0.609 & -4.274 $\pm$ 0.343 & 23.4 & $-$2.63 $\pm$ 0.21 & -2.65 $\pm$ 0.09 & 0.00 &  N  \\ \\ 
298.08961667 & $-$22.02933056 & 19.42 $\pm$ 0.01 & 18.67 $\pm$ 0.01 & 20.61 $\pm$ 0.02 & 80.2 $\pm$ 1.9 & -1.374 $\pm$ 0.597 & -5.162 $\pm$ 0.334 & 23.2 & $-$1.03 $\pm$ 0.12 & --- & 0.71 &  N  \\ \\ 
298.04697083 & $-$22.09121389 & 19.76 $\pm$ 0.01 & 19.18 $\pm$ 0.01 & 20.34 $\pm$ 0.02 & 79.1 $\pm$ 2.4 & -17.5 $\pm$ 0.76 & -1.088 $\pm$ 0.455 & 22.5 & $-$2.88 $\pm$ 0.24 & -2.8 $\pm$ 0.16 & 0.00 &  N  \\ \\ 
298.07480417 & $-$22.08419444 & 20.88 $\pm$ 0.01 & 19.95 $\pm$ 0.01 & 21.93 $\pm$ 0.05 & 55.5 $\pm$ 3.3 & -10.032 $\pm$ 1.951 & -18.831 $\pm$ 1.084 & 14.2 & $-$2.62 $\pm$ 0.21 & --- & 0.00 &  N  \\ \\ 
298.05488333 & $-$22.04129444 & 20.87 $\pm$ 0.01 & 20.23 $\pm$ 0.01 & 21.93 $\pm$ 0.05 & 21.2 $\pm$ 5.4 & 1.271 $\pm$ 4.225 & -12.49 $\pm$ 2.044 & 9.7 & --- & --- & 0.64 &  N  \\ \\ 
298.10188750 & $-$22.06900000 & 20.94 $\pm$ 0.01 & 20.45 $\pm$ 0.01 & 21.52 $\pm$ 0.04 & -55.2 $\pm$ 4.9 & --- & --- & 9.2 & $-$2.27 $\pm$ 0.28 & --- & 0.00 &  N  \\ \\ 
298.03362500 & $-$22.09989444 & 20.97 $\pm$ 0.01 & 20.35 $\pm$ 0.01 & 21.89 $\pm$ 0.05 & -85.2 $\pm$ 5.5 & 0.485 $\pm$ 3.664 & -2.08 $\pm$ 2.111 & 15.5 & $-$1.34 $\pm$ 0.21 & --- & 0.35 &  N  \\ \\ 
298.05201667 & $-$22.07245833 & 21.11 $\pm$ 0.01 & 20.76 $\pm$ 0.01 & 21.48 $\pm$ 0.04 & -270.8 $\pm$ 7.6 & --- & --- & 7.0 & $-$3.08 $\pm$ 0.45 & --- & 0.00 &  N  \\ \\ 
298.07867917 & $-$22.07232778 & 21.36 $\pm$ 0.01 & 20.79 $\pm$ 0.01 & 22.19 $\pm$ 0.06 & 21.9 $\pm$ 7.0 & --- & --- & 6.5 & $-$1.53 $\pm$ 0.26 & --- & 0.91 &  N  \\ \\ 
298.06682917 & $-$22.03141944 & 21.36 $\pm$ 0.01 & 20.58 $\pm$ 0.01 & 22.71 $\pm$ 0.09 & -36.9 $\pm$ 6.0 & --- & --- & 7.0 & --- & --- & 0.00 &  N  \\ \\ 
298.03616667 & $-$22.06642778 & 21.75 $\pm$ 0.01 & 21.09 $\pm$ 0.01 & 22.78 $\pm$ 0.10 & 162.8 $\pm$ 6.8 & --- & --- & 4.6 & $-$1.21 $\pm$ 0.45 & --- & 0.01 &  N  \\ \\ 
298.02500000 & $-$22.06503611 & 22.08 $\pm$ 0.01 & 21.27 $\pm$ 0.02 & 23.50 $\pm$ 0.18 & 57.2 $\pm$ 5.8 & --- & --- & 3.7 & --- & --- & 0.00 &  N  \\ \\ 
298.05970000 & $-$22.05247500 & 22.40 $\pm$ 0.02 & 21.46 $\pm$ 0.02 & $-$0.460 $\pm$ 0.00 & 27.9 $\pm$ 6.5 & --- & --- & 4.1 & --- & --- & 0.00 &  N  \\ \\ 
298.04913330 & $-$22.03305435 & 17.50 $\pm$ 0.01 & 16.59 $\pm$ 0.01 & 19.05 $\pm$ 0.01 & 56.1 $\pm$ 1.6 & 4.746 $\pm$ 0.162 & 0.185 $\pm$ 0.092 & 40.1 & --- & -1.3 $\pm$ 0.09 & 0.00 &  N  \\ \\ 
298.07669067 & $-$22.08716202 & 17.29 $\pm$ 0.01 & 15.89 $\pm$ 0.01 & 19.34 $\pm$ 0.01 & -37.0 $\pm$ 1.5 & -1.479 $\pm$ 0.125 & -5.347 $\pm$ 0.069 & 44.1 & --- & -1.18 $\pm$ 0.08 & 0.00 &  N  \\ \\ 
298.21893311 & $-$22.07197189 & 17.44 $\pm$ 0.01 & 16.54 $\pm$ 0.01 & 19.05 $\pm$ 0.01 & -6.6 $\pm$ 1.7 & 3.183 $\pm$ 0.205 & -7.464 $\pm$ 0.116 & 37.3 & --- & -1.18 $\pm$ 0.09 & 0.00 &  N  \\ \\ 

\end{tabular}
\end{sideways}
\end{table*}

\begin{table*}
\renewcommand\thetable{3} 
\caption{Velocities and individual metallicities for all stars observed more than once, per mask. Mask 1 was observed on the 2015-09-18, mask 2 on the 2015-09-08 and mask 3 on the 2015-09-12 (respectively 2457283.760868, 2457273.738102 and 2457277.742083 in Julian dates). The systematic threshold $\delta_{\mathrm{thr}}$ is not included in the velocity uncertainties presented in this table. The individual spectroscopic metallicity is reported for stars with S/N $>= 12$ and $g_0 > 20.5$ only.
\label{tbl-2}}
\setlength{\tabcolsep}{2.5pt}
\renewcommand{\arraystretch}{0.3}
\begin{tabular}{cccccccccccccc}
\hline

RA (deg) & DEC (deg) & Mask & $v_{r} (\kms)$ & [Fe/H]$_\mathrm{spectro}$  \\
\hline

$298.18001250$ & $$-$22.071750000$ & Combined & $0.2 \pm 0.9$ & --- \\ \\ 
 &  & Mask 1 & $0.2 \pm 1.6$ & --- \\ \\ 
 &  & Mask 2 & $1.1 \pm 1.7$ & --- \\ \\ 
 &  & Mask 3 & $-0.5 \pm 1.4$ & --- \\ \\ 
$298.16145833$ & $$-$22.082655560$ & Combined & $-182.8 \pm 0.9$ & $-2.27 \pm 0.04$ \\ \\ 
 &  & Mask 1 & $-182.4 \pm 1.8$ & $-2.36 \pm 0.07$ \\ \\ 
 &  & Mask 2 & $-183.1 \pm 1.4$ & $-2.25 \pm 0.08$ \\ \\ 
 &  & Mask 3 & $-182.8 \pm 1.4$ & $-2.2 \pm 0.07$ \\ \\ 
$298.18212917$ & $$-$22.057094440$ & Combined & $28.4 \pm 1.8$ & --- \\ \\ 
 &  & Mask 1 & $29.4 \pm 5.6$ & --- \\ \\ 
 &  & Mask 2 & $29.6 \pm 2.4$ & --- \\ \\ 
 &  & Mask 3 & $25.7 \pm 3.3$ & --- \\ \\ 
$298.20599167$ & $$-$21.987900000$ & Combined & $16.6 \pm 1.3$ & --- \\ \\ 
 &  & Mask 1 & $17.4 \pm 1.7$ & --- \\ \\ 
 &  & Mask 3 & $15.1 \pm 2.2$ & --- \\ \\ 
$298.13157500$ & $$-$21.985827780$ & Combined & $27.5 \pm 1.5$ & --- \\ \\ 
 &  & Mask 1 & $27.1 \pm 2.0$ & --- \\ \\ 
 &  & Mask 3 & $28.0 \pm 2.4$ & --- \\ \\ 
$298.13137917$ & $$-$21.982733330$ & Combined & $-39.4 \pm 2.2$ & --- \\ \\ 
 &  & Mask 1 & $-44.0 \pm 2.8$ & --- \\ \\ 
 &  & Mask 3 & $-31.4 \pm 3.6$ & --- \\ \\ 
$298.15507917$ & $$-$21.980541670$ & Combined & $12.5 \pm 1.7$ & --- \\ \\ 
 &  & Mask 1 & $12.0 \pm 2.9$ & --- \\ \\ 
 &  & Mask 3 & $12.7 \pm 2.1$ & --- \\ \\ 
$298.15097500$ & $$-$21.952833330$ & Combined & $140.9 \pm 1.7$ & --- \\ \\ 
 &  & Mask 1 & $140.8 \pm 2.1$ & --- \\ \\ 
 &  & Mask 3 & $141.1 \pm 3.1$ & --- \\ \\ 
$298.19072083$ & $$-$21.967580560$ & Combined & $-118.1 \pm 6.2$ & $-2.86 \pm 0.23$ \\ \\ 
 &  & Mask 1 & $-116.0 \pm 10.1$ & $-3.02 \pm 0.27$ \\ \\ 
 &  & Mask 3 & $-119.4 \pm 7.9$ & $-2.46 \pm 0.42$ \\ \\ 
$298.18688333$ & $$-$21.977138890$ & Combined & $-107.3 \pm 2.8$ & $-1.26 \pm 0.13$ \\ \\ 
 &  & Mask 1 & $-114.2 \pm 5.4$ & $-1.31 \pm 0.17$ \\ \\ 
 &  & Mask 3 & $-104.7 \pm 3.3$ & $-1.21 \pm 0.18$ \\ \\ 
$298.18320417$ & $$-$21.963619440$ & Combined & $-15.4 \pm 3.5$ & --- \\ \\ 
 &  & Mask 1 & $-9.8 \pm 7.3$ & --- \\ \\ 
 &  & Mask 3 & $-17.1 \pm 4.0$ & --- \\ \\ 
$298.13665417$ & $$-$21.974238890$ & Combined & $-7.8 \pm 7.4$ & --- \\ \\ 
 &  & Mask 1 & $-1.4 \pm 9.9$ & --- \\ \\ 
 &  & Mask 3 & $-16.3 \pm 11.3$ & --- \\ \\ 
$298.19440000$ & $$-$21.990705560$ & Combined & $-67.3 \pm 3.8$ & --- \\ \\ 
 &  & Mask 1 & $-64.4 \pm 7.0$ & --- \\ \\ 
 &  & Mask 3 & $-68.5 \pm 4.5$ & --- \\ \\ 
$298.18154167$ & $$-$21.969627780$ & Combined & $30.1 \pm 16.0$ & --- \\ \\ 
 &  & Mask 1 & $26.3 \pm 20.0$ & --- \\ \\ 
 &  & Mask 3 & $37.0 \pm 26.9$ & --- \\ \\ 
$298.19622500$ & $$-$21.993688890$ & Combined & $-2.0 \pm 13.4$ & --- \\ \\ 
 &  & Mask 1 & $-4.6 \pm 17.6$ & --- \\ \\ 
 &  & Mask 3 & $1.6 \pm 20.5$ & --- \\ \\ 
$298.16959583$ & $$-$22.174630560$ & Combined & $89.7 \pm 1.1$ & --- \\ \\ 
 &  & Mask 1 & $88.5 \pm 1.7$ & --- \\ \\ 
 &  & Mask 3 & $90.5 \pm 1.5$ & --- \\ \\ 
$298.16810000$ & $$-$22.188300000$ & Combined & $105.1 \pm 1.3$ & --- \\ \\ 
 &  & Mask 1 & $104.8 \pm 1.5$ & --- \\ \\ 
 &  & Mask 3 & $105.9 \pm 2.5$ & --- \\ \\ 
$298.19661667$ & $$-$22.146352780$ & Combined & $29.2 \pm 1.8$ & --- \\ \\ 
 &  & Mask 1 & $30.8 \pm 2.4$ & --- \\ \\ 
 &  & Mask 3 & $27.1 \pm 2.8$ & --- \\ \\ 
$298.17483750$ & $$-$22.166063890$ & Combined & $-54.0 \pm 1.3$ & --- \\ \\ 
 &  & Mask 1 & $-54.2 \pm 1.9$ & --- \\ \\ 
 &  & Mask 3 & $-53.9 \pm 1.7$ & --- \\ \\ 
$298.19827500$ & $$-$22.144983330$ & Combined & $89.3 \pm 1.3$ & --- \\ \\ 
 &  & Mask 1 & $91.0 \pm 2.4$ & --- \\ \\ 
 &  & Mask 3 & $88.5 \pm 1.6$ & --- \\ \\ 
$298.16120000$ & $$-$22.008286110$ & Combined & $-85.1 \pm 1.4$ & --- \\ \\ 
 &  & Mask 1 & $-82.0 \pm 2.1$ & --- \\ \\ 
 &  & Mask 3 & $-87.5 \pm 1.9$ & --- \\ \\

\end{tabular}
\end{table*}

\begin{table*}
\renewcommand\thetable{3} 
\caption{Velocities and individual metallicities for all stars observed more than once, per mask. - Part 2
\label{tbl-2}}
\setlength{\tabcolsep}{2.5pt}
\renewcommand{\arraystretch}{0.3}
\begin{tabular}{cccccccccccccc}
\hline

$298.15403750$ & $$-$22.111083330$ & Combined & $51.4 \pm 1.9$ & --- \\ \\ 
 &  & Mask 1 & $51.2 \pm 3.3$ & --- \\ \\ 
 &  & Mask 3 & $51.5 \pm 2.4$ & --- \\ \\ 
$298.15852917$ & $$-$22.058469440$ & Combined & $-167.9 \pm 9.5$ & $-0.44 \pm 0.28$ \\ \\ 
 &  & Mask 1 & $-161.0 \pm 23.3$ & $-1.27 \pm 0.77$ \\ \\ 
 &  & Mask 3 & $-169.3 \pm 10.4$ & $-0.32 \pm 0.3$ \\ \\ 
$298.19296667$ & $$-$22.022188890$ & Combined & $122.6 \pm 1.8$ & --- \\ \\ 
 &  & Mask 1 & $123.3 \pm 2.8$ & --- \\ \\ 
 &  & Mask 3 & $122.0 \pm 2.5$ & --- \\ \\ 
$298.17763750$ & $$-$22.046011110$ & Combined & $-135.3 \pm 4.2$ & $-0.93 \pm 0.28$ \\ \\ 
 &  & Mask 1 & $-177.0 \pm 4.9$ & $-1.25 \pm 0.39$ \\ \\ 
 &  & Mask 3 & $-12.9 \pm 8.4$ & $-0.61 \pm 0.39$ \\ \\ 
$298.17252500$ & $$-$22.074108330$ & Combined & $-76.1 \pm 10.3$ & $-0.07 \pm 0.23$ \\ \\ 
 &  & Mask 1 & $-82.6 \pm 18.3$ & $-0.07 \pm 0.35$ \\ \\ 
 &  & Mask 3 & $-73.1 \pm 12.5$ & $-0.07 \pm 0.31$ \\ \\ 
$298.12763750$ & $$-$22.172886110$ & Combined & $-15.5 \pm 2.6$ & $-1.78 \pm 0.16$ \\ \\ 
 &  & Mask 1 & $-13.7 \pm 3.4$ & $-1.78 \pm 0.2$ \\ \\ 
 &  & Mask 3 & $-17.9 \pm 4.0$ & $-1.78 \pm 0.25$ \\ \\ 
$298.20524167$ & $$-$22.027505560$ & Combined & $201.1 \pm 3.4$ & --- \\ \\ 
 &  & Mask 1 & $212.3 \pm 4.9$ & --- \\ \\ 
 &  & Mask 3 & $190.9 \pm 4.7$ & --- \\ \\ 
$298.14820833$ & $$-$22.002480560$ & Combined & $163.6 \pm 1.9$ & $-2.51 \pm 0.12$ \\ \\ 
 &  & Mask 1 & $163.3 \pm 2.4$ & $-2.43 \pm 0.16$ \\ \\ 
 &  & Mask 3 & $164.1 \pm 3.4$ & $-2.62 \pm 0.18$ \\ \\ 
$298.16396667$ & $$-$22.063497220$ & Combined & $41.4 \pm 2.0$ & $-2.02 \pm 0.11$ \\ \\ 
 &  & Mask 1 & $42.2 \pm 3.0$ & $-2.01 \pm 0.16$ \\ \\ 
 &  & Mask 3 & $40.7 \pm 2.8$ & $-2.02 \pm 0.16$ \\ \\ 
$298.16217500$ & $$-$22.054411110$ & Combined & $-176.0 \pm 1.5$ & $-2.24 \pm 0.08$ \\ \\ 
 &  & Mask 1 & $-176.4 \pm 1.8$ & $-2.2 \pm 0.12$ \\ \\ 
 &  & Mask 3 & $-175.1 \pm 2.7$ & $-2.26 \pm 0.1$ \\ \\ 
$298.14762500$ & $$-$22.189836110$ & Combined & $-227.3 \pm 3.3$ & --- \\ \\ 
 &  & Mask 1 & $-229.5 \pm 7.4$ & --- \\ \\ 
 &  & Mask 3 & $-226.8 \pm 3.7$ & --- \\ \\ 
$298.19400833$ & $$-$22.086383330$ & Combined & $-174.9 \pm 2.3$ & $-2.1 \pm 0.11$ \\ \\ 
 &  & Mask 1 & $-174.8 \pm 2.7$ & $-2.11 \pm 0.15$ \\ \\ 
 &  & Mask 3 & $-175.1 \pm 4.4$ & $-2.08 \pm 0.15$ \\ \\ 
$298.19723333$ & $$-$22.124525000$ & Combined & $-323.9 \pm 2.6$ & --- \\ \\ 
 &  & Mask 1 & $-320.8 \pm 3.7$ & --- \\ \\ 
 &  & Mask 3 & $-326.8 \pm 3.6$ & --- \\ \\ 
$298.14902500$ & $$-$22.023586110$ & Combined & $-71.1 \pm 3.6$ & --- \\ \\ 
 &  & Mask 1 & $-68.0 \pm 4.1$ & --- \\ \\ 
 &  & Mask 3 & $-80.9 \pm 7.3$ & --- \\ \\ 
$298.18112083$ & $$-$22.060772220$ & Combined & $-179.5 \pm 2.4$ & --- \\ \\ 
 &  & Mask 1 & $-179.9 \pm 3.4$ & --- \\ \\ 
 &  & Mask 3 & $-179.2 \pm 3.5$ & --- \\ \\ 
$298.17317500$ & $$-$22.115836110$ & Combined & $-173.9 \pm 3.8$ & --- \\ \\ 
 &  & Mask 1 & $-157.7 \pm 6.4$ & --- \\ \\ 
 &  & Mask 3 & $-182.8 \pm 4.7$ & --- \\ \\ 
$298.12500000$ & $$-$22.117663890$ & Combined & $-179.7 \pm 16.8$ & --- \\ \\ 
 &  & Mask 1 & $-97.2 \pm 42.3$ & --- \\ \\ 
 &  & Mask 3 & $-195.1 \pm 18.3$ & --- \\ \\ 
$298.14954583$ & $$-$22.107030560$ & Combined & $-177.4 \pm 3.7$ & --- \\ \\ 
 &  & Mask 1 & $-180.8 \pm 6.1$ & --- \\ \\ 
 &  & Mask 3 & $-175.5 \pm 4.7$ & --- \\ \\ 
$298.18245417$ & $$-$22.105638890$ & Combined & $-176.3 \pm 6.0$ & --- \\ \\ 
 &  & Mask 1 & $-172.9 \pm 11.1$ & --- \\ \\ 
 &  & Mask 3 & $-177.7 \pm 7.1$ & --- \\ \\ 
$298.16334583$ & $$-$22.143216670$ & Combined & $71.0 \pm 4.9$ & --- \\ \\ 
 &  & Mask 1 & $75.7 \pm 6.8$ & --- \\ \\ 
 &  & Mask 3 & $65.8 \pm 7.0$ & --- \\ \\ 
$298.16188333$ & $$-$22.053213890$ & Combined & $-177.5 \pm 7.8$ & --- \\ \\ 
 &  & Mask 1 & $-178.1 \pm 12.5$ & --- \\ \\ 
 &  & Mask 3 & $-177.1 \pm 9.9$ & --- \\ \\ 
$298.18229167$ & $$-$22.042752780$ & Combined & $-177.1 \pm 3.9$ & --- \\ \\ 
 &  & Mask 1 & $-180.6 \pm 6.3$ & --- \\ \\ 
 &  & Mask 3 & $-175.0 \pm 5.0$ & --- \\ \\ 
$298.19592917$ & $$-$22.132950000$ & Combined & $22.0 \pm 6.6$ & --- \\ \\ 
 &  & Mask 1 & $-497.1 \pm 14.1$ & --- \\ \\ 
 &  & Mask 3 & $166.5 \pm 7.4$ & --- \\ \\

\end{tabular}
\end{table*}

\begin{table*}
\renewcommand\thetable{3} 
\caption{Velocities and individual metallicities for all stars observed more than once, per mask. - Part 3
\label{tbl-2}}
\setlength{\tabcolsep}{2.5pt}
\renewcommand{\arraystretch}{0.3}
\begin{tabular}{cccccccccccccc}
\hline

$298.13948750$ & $$-$22.176200000$ & Combined & $174.5 \pm 11.6$ & --- \\ \\ 
 &  & Mask 1 & $142.2 \pm 24.7$ & --- \\ \\ 
 &  & Mask 3 & $183.7 \pm 13.2$ & --- \\ \\ 
$298.15003333$ & $$-$22.017416670$ & Combined & $-276.5 \pm 3.6$ & --- \\ \\ 
 &  & Mask 1 & $-273.9 \pm 7.0$ & --- \\ \\ 
 &  & Mask 3 & $-277.5 \pm 4.2$ & --- \\ \\ 
$298.14941250$ & $$-$22.179366670$ & Combined & $-59.4 \pm 5.6$ & --- \\ \\ 
 &  & Mask 1 & $-52.7 \pm 6.8$ & --- \\ \\ 
 &  & Mask 3 & $-73.5 \pm 9.8$ & --- \\ \\ 
$298.16383333$ & $$-$22.186166670$ & Combined & $-11.5 \pm 5.2$ & --- \\ \\ 
 &  & Mask 1 & $-6.4 \pm 6.4$ & --- \\ \\ 
 &  & Mask 3 & $-21.4 \pm 9.0$ & --- \\ \\ 
$298.16461667$ & $$-$22.091652780$ & Combined & $-172.1 \pm 4.5$ & --- \\ \\ 
 &  & Mask 1 & $-164.5 \pm 5.5$ & --- \\ \\ 
 &  & Mask 3 & $-189.6 \pm 8.2$ & --- \\ \\ 
$298.20289583$ & $$-$22.035700000$ & Combined & $-177.4 \pm 7.9$ & --- \\ \\ 
 &  & Mask 1 & $-181.2 \pm 10.3$ & --- \\ \\ 
 &  & Mask 3 & $-171.9 \pm 12.3$ & --- \\ \\ 
$298.17171250$ & $$-$22.122622220$ & Combined & $-109.2 \pm 9.5$ & --- \\ \\ 
 &  & Mask 1 & $-112.2 \pm 17.1$ & --- \\ \\ 
 &  & Mask 3 & $-107.9 \pm 11.4$ & --- \\ \\ 
$298.16051250$ & $$-$22.109411110$ & Combined & $-51.8 \pm 8.5$ & --- \\ \\ 
 &  & Mask 1 & $-12.1 \pm 9.9$ & --- \\ \\ 
 &  & Mask 3 & $-171.3 \pm 17.1$ & --- \\ \\ 
$298.16816250$ & $$-$22.184716670$ & Combined & $62.8 \pm 14.7$ & --- \\ \\ 
 &  & Mask 1 & $48.7 \pm 16.5$ & --- \\ \\ 
 &  & Mask 3 & $116.8 \pm 32.3$ & --- \\ \\ 
$298.17825417$ & $$-$22.015291670$ & Combined & $341.2 \pm 10.5$ & --- \\ \\ 
 &  & Mask 1 & $375.7 \pm 10.9$ & --- \\ \\ 
 &  & Mask 3 & $-85.7 \pm 38.3$ & --- \\ \\ 
$298.20325417$ & $$-$22.119213890$ & Combined & $268.2 \pm 8.4$ & --- \\ \\ 
 &  & Mask 1 & $737.8 \pm 29.0$ & --- \\ \\ 
 &  & Mask 3 & $225.0 \pm 8.8$ & --- \\ \\ 
$298.12890833$ & $$-$22.157383330$ & Combined & $31.0 \pm 6.0$ & --- \\ \\ 
 &  & Mask 1 & $32.6 \pm 7.5$ & --- \\ \\ 
 &  & Mask 3 & $28.1 \pm 9.9$ & --- \\ \\ 
$298.18128333$ & $$-$22.114341670$ & Combined & $-15.4 \pm 10.0$ & --- \\ \\ 
 &  & Mask 1 & $-15.8 \pm 14.1$ & --- \\ \\ 
 &  & Mask 3 & $-15.0 \pm 14.2$ & --- \\ \\ 
$298.12620417$ & $$-$22.163083330$ & Combined & $-100.2 \pm 7.3$ & --- \\ \\ 
 &  & Mask 1 & $-92.8 \pm 7.9$ & --- \\ \\ 
 &  & Mask 3 & $-141.3 \pm 18.6$ & --- \\ \\ 
$298.19941250$ & $$-$22.102163890$ & Combined & $332.5 \pm 8.3$ & --- \\ \\ 
 &  & Mask 1 & $-238.7 \pm 19.6$ & --- \\ \\ 
 &  & Mask 3 & $457.9 \pm 9.2$ & --- \\ \\ 
$298.13974583$ & $$-$22.025369440$ & Combined & $642.6 \pm 13.9$ & --- \\ \\ 
 &  & Mask 1 & $574.9 \pm 20.1$ & --- \\ \\ 
 &  & Mask 3 & $704.6 \pm 19.2$ & --- \\ \\ 
$298.17350417$ & $$-$22.098127780$ & Combined & $37.3 \pm 8.2$ & --- \\ \\ 
 &  & Mask 1 & $724.5 \pm 15.4$ & --- \\ \\ 
 &  & Mask 3 & $-238.5 \pm 9.8$ & --- \\ \\ 
$298.19098333$ & $$-$22.088255560$ & Combined & $-305.3 \pm 16.6$ & --- \\ \\ 
 &  & Mask 1 & $-622.7 \pm 20.8$ & --- \\ \\ 
 &  & Mask 3 & $257.3 \pm 27.6$ & --- \\ \\ 
$298.14785000$ & $$-$22.050347220$ & Combined & $461.1 \pm 10.1$ & --- \\ \\ 
 &  & Mask 1 & $537.3 \pm 26.9$ & --- \\ \\ 
 &  & Mask 3 & $448.6 \pm 10.9$ & --- \\ \\ 
$298.15192083$ & $$-$22.029111110$ & Combined & $126.0 \pm 30.7$ & --- \\ \\ 
 &  & Mask 1 & $139.2 \pm 43.3$ & --- \\ \\ 
 &  & Mask 3 & $112.6 \pm 43.6$ & --- \\ \\ 
$298.16238403$ & $$-$22.077482220$ & Combined & $-170.4 \pm 0.7$ & $-2.1 \pm 0.04$ \\ \\ 
 &  & Mask 1 & $-173.2 \pm 1.2$ & $-2.14 \pm 0.06$ \\ \\ 
 &  & Mask 3 & $-169.0 \pm 0.9$ & $-2.05 \pm 0.06$ \\ \\ 
$298.16424561$ & $$-$22.168029780$ & Combined & $-82.8 \pm 1.1$ & $-1.39 \pm 0.06$ \\ \\ 
 &  & Mask 1 & $-83.0 \pm 1.7$ & $-1.44 \pm 0.08$ \\ \\ 
 &  & Mask 3 & $-82.6 \pm 1.5$ & $-1.33 \pm 0.08$ \\ \\

 \end{tabular}
\end{table*}

\newcommand{\mnras}{MNRAS}
\newcommand{\pasa}{PASA}
\newcommand{\nat}{Nature}
\newcommand{\araa}{ARAA}
\newcommand{\aj}{AJ}
\newcommand{\apj}{ApJ}
\newcommand{\apjl}{ApJ}
\newcommand{\apjs}{ApJSupp}
\newcommand{\aap}{A\&A}
\newcommand{\aaps}{A\&ASupp}
\newcommand{\pasp}{PASP}


\clearpage

\end{document}